\documentclass[a4paper,preprintnumbers,floatfix,superscriptaddress,pra,twocolumn,showpacs,notitlepage,longbibliography]{revtex4-2} 

\usepackage[utf8]{inputenc}  %Input what you want e.g., é, ł, a, ü
\usepackage[T1]{fontenc}     %Output what you want e.g., é, ł, a, ü
\usepackage[english]{babel}  %Do hyphenation according to British english

\usepackage[colorlinks,citecolor=myBlue,linkcolor=magenta,urlcolor=myBlue]{hyperref}  %Hyperlinks (pink, green, blue)
\usepackage{graphicx} % Package to insert external figures
\usepackage{tikz}
\usepackage{amssymb,amsthm,mathtools,dsfont}
\usepackage{physics}
\usepackage[capitalise]{cleveref}			
\usepackage{units} %fuer nicefrac
\usepackage[normalem]{ulem} % Add commands like \sout
\usepackage{gensymb} %fuer \degree
\usepackage{todonotes} %Todos einfuegen mit \todo{text}

\usetikzlibrary{calc,positioning,backgrounds,fit}
\usetikzlibrary{arrows.meta}

\usepackage{soul}

\newtheorem{observation}{Observation}
\newtheorem{theorem}[observation]{Theorem}
\newtheorem{definition}[observation]{Definition}
\newtheorem{remark}[observation]{Remark}

\crefname{theorem}{Thm.}{Thms.}
\crefname{table}{Tbl.}{Tbls.}
\crefname{appendix}{App.}{Apps.}
\crefname{section}{Sec.}{Secs.}
\crefname{remark}{Rem.}{Rems.}

\DeclareMathOperator{\CC}{\mathcal{C}}
\DeclareMathOperator{\GG}{\mathcal{G}}
\DeclareMathOperator{\HH}{\mathcal{H}}
\DeclareMathOperator{\OO}{\mathcal{O}}
\DeclareMathOperator{\QQ}{\mathcal{Q}}

\DeclareMathOperator{\1}{\mathds{1}}
\DeclareMathOperator{\Rd}{\mathds{R}}

\DeclareMathOperator{\sq}{\sqrt{2}}

\DeclareMathOperator{\Schsh}{S_{\textup{CHSH}}}

\DeclareMathOperator{\Qchsh}{\QQ_{\textup{CHSH}}}
\DeclareMathOperator{\Qshad}{Q_{\textup{CSW}}}
\DeclareMathOperator{\Qdrei}{\QQ_{33,33}}
\DeclareMathOperator{\Qvier}{\QQ_{44,1111}}
\DeclareMathOperator{\Gchsh}{\GG_{\textup{CHSH}}}
\DeclareMathOperator{\Gshad}{G_{\textup{CSW}}}
\DeclareMathOperator{\Gdrei}{\GG_{33,33}}
\DeclareMathOperator{\Gvier}{\GG_{44,1111}}

\DeclareMathOperator{\Tshad}{TH(\Gshad)}
\DeclareMathOperator{\Tg}{TH_c(\GG)}
\DeclareMathOperator{\Th}{TH}

\definecolor{myRed}{HTML}{F39200}
\definecolor{myBlue}{HTML}{003361}
\definecolor{myBlue2}{HTML}{004e9b}

\begin{document}
%%%%%%%%%%%%%%%%%%%%%%%%%%%%%%%%%%%%%%%%%%%%%%%%%%%%%%%%%

\title{Quantum sets of the multicolored-graph approach to contextuality} 

\author{Lina Vandr\'e}
\affiliation{Naturwissenschaftlich-Technische Fakultät, Universität Siegen, Walter-Flex-Straße 3, 57068 Siegen, Germany}
\affiliation{Institute for Theoretical Physics, Technikerstra\ss e 21a, 6020 Innsbruck, Austria}
\affiliation{Departamento de Matemática Aplicada, Instituto de Matemática, Estatística e Computação Cient\'ifica, Universidade Estadual de Campinas (Unicamp),  13083-859,  Campinas,  São Paulo,  Brazil}
\author{Marcelo {Terra Cunha}}
\affiliation{Departamento de Matemática Aplicada, Instituto de Matemática, Estatística e Computação Cient\'ifica, Universidade Estadual de Campinas (Unicamp),  13083-859,  Campinas,  São Paulo,  Brazil}

\date{\today}

\begin{abstract}

The Clauser-Horne-Shimony-Holt (CHSH) inequalities are the most famous examples of Bell inequalities. Cabello, Severini, and Winter  came up with a graph approach to noncontextuality inequalities, which connects some graph-theoretic concepts to quantum and classical correlations. 
For example, the theta body of the exclusivity graph can be associated with the set of correlations achieved by quantum theory. Following the Cabello-Severini-Winter (CSW) approach, one may think that the theta body of the CHSH graph, 
$\text{TH}(G_{\text{CHSH}})$, is equal to the quantum set of the CHSH Bell inequality $\mathcal{Q}_{\textup{CHSH}}$, but is this really true? All assumptions about the CHSH inequalities come from Bell scenarios, while CSW approach only demands the exclusivity structure of a non-contextuality (NC) scenario. To deal with the extra structure related to the presence of different players in a Bell scenario like CHSH, the colored-graph approach was introduced. Does it make any difference to think about CHSH as a Bell scenario or a more general NC scenario? 
 The Bell CHSH inequality is represented by a bicolored graph $\mathcal{G}$ and the NC CHSH inequality by a simple graph $G$, which is the shadow of the colored graph $\mathcal{G}$. 
In general, we have that the theta body of the colored graph $\text{TH}_c(\mathcal{G})$ is a subset of the theta body of its shadow graph 
$\text{TH}(G)$ in the same way that the Lov\'asz number, which corresponds to the quantum bound, of the simple graph $\vartheta(G)$ is greater than or equal to the Lov\'asz number of the colored graph $ \vartheta_c(\mathcal{G})$. 
In the case of the CHSH inequality, we have that $\vartheta(G) = \vartheta_c(\mathcal{G})$. Does this accident also hold for the corresponding quantum sets?
 Is it true that $\text{TH}(G) = \text{TH}_c(\mathcal{G})$, which would mean that every correlation reached by quantum theory applied to the CHSH NC scenario could also be obtained at the in principle more restrictive CHSH Bell scenario? 
In this paper our answer to such a question is negativ. 
We show that $\text{TH}_c(\mathcal{G}) \subsetneq \Th(G)$ and therefore that there are quantum correlations which can not be obtained under Bell restrictions.

\end{abstract}

\maketitle

\section{Introduction}

Bell showed that quantum theory can not be explained with non-contextual hidden variables or with local hidden variables \cite{Bell66, Bell}.
One way to show the impossibility of describing quantum theory with hidden variable theories is with Bell inequalities or the more general non-contextuality (NC) inequalities.
The most popular example of a Bell inequality is the Clauser-Horne-Shimony-Holt (CHSH) Bell inequality \cite{CHSH}. 
The simplest non-contextuality inequality which is not a Bell inequality is the Klyachko-Can-Binicioğlu-Shumovsky (KCBS) inequality \cite{KCBS}.
Pentagonal Bell inequalities \cite{PentBI} were later introduced, showing that the graph structure could also support Bell inequalities.
Pentagonal Bell inequalities \cite{PentBI} give good examples of different Bell inequalities sharing the same  exclusivity graph, but smaller quantum bounds, with an an NC inequality.

Bell scenarios consist of several parties. The parties share a common state and perform  local measurements on their parts of the system. 
The measurements of different parties are compatible, while alternative (incompatible) measurements are available to each part. 
Specific Bell experiments prescribe the number of parties, the number of measurements among which they can choose, and the number of elements in the outcome set.
Bell inequalities have the restriction that compatible measurements are performed on spatially separated subsystems. 
For general non-contextuality inequalities we do not necessarily need the notion of parties. 
Instead we deal with contexts: sets of jointly compatible measurements. 
This measurements can be performed in a single laboratory or by different parties in spacelike separated laboratories. 
A Bell or a NC inequality can be written as a positive linear combination $S$ of probabilities of events and a number $\alpha$ which is {an upper bound to} classically reachable values within the given inequality (details in \cref{Cgraphappr}). 

The exclusivity relations of events can be represented by exclusivity graphs. 
The relation between NC inequalities and graph-theoretic concepts applied to exclusivity graphs was first noted by Cabello, Severini, and Winter  \cite{CSW1, CSW2}.
In exclusivity graphs, events get represented by vertices and exclusivity between events by edges. 
This graph encodes the central ingredients of the NC inequality and graph theory brings clarity to many optimization processes, for example, while the independence number of the graph gives the  classical bound, the Lov\'asz number is closely related to the quantum bound \cite{Lovasz,Tsirelson, sandwich}.

In multipartite scenarios, i.e., scenarios where locality plays a role, we can further differentiate among exclusivities originated in different parties and even multiple exclusivities, coming jointly from different parts. 
To include this differentiation into the notion of exclusivity graphs, 
Rabelo \textit{et al.}\ extended the approach to edge-colored graphs, where 
colors are used to represent the parts \cite{MultGraphAppr}.
In a colored exclusivity graph, two vertices are connected by an edge of a specific color if that exclusivity comes from that specific part.
Multiple edges become justifiable when two events are exclusive for more than one part. 
The multicolored exclusivity graph represents exclusivity between events and also points out which parts make them exclusive.
This approach, first introduced to deal with the extra restrictions imposed by Bell scenario, describes the correlation structure of multipartite (but not necessarily Bell) inequalities more precisely.
A more restrictive graph invariant, the colored Lov\'asz number, can capture the extra requirements imposed by locality, allowing for tight bounds where the usual Lov\'asz number could only be an upper bound.
When parties are involved, the simple graph coming from the Cabello-Severini-Winter (CSW) approach is called the shadow of the colored graph.

There are examples, such as the pentagonal Bell inequalities, where the colored graph approach leads to a tight upper bound of the maximum quantum value, while the simple graph does not.  
Nevertheless, in the case of the CHSH inequality, the so far analyzed properties of the colored and simple graphs are identical.
This raises the  question of whether for the CHSH inequality  it is possible that the simple graph already encodes all the restrictions.
By comparing the quantum sets of the simple and colored graph, we show that the colored graph is indeed a more restrictive description.
In other words, there are behaviors which can be achieved within quantum theory under CHSH NC restrictions, but not under Bell restrictions.
To do so, we construct a family of graphs which have intermediate quantum sets. 
The graphs of this family are subgraphs of the colored CHSH graph which all have the same {shadow}. 
Since the quantum set of a graph is convex, we explore it by looking at the maximal value that a linear function can
take on this quantum set for various linear functions.

This paper confirms that for the CHSH structure, the colored graph is not only a more precise but also a more strict description of Bell conditions.
Since the argument originated from a CHSH family of colored graphs with the same shadow, this adds to the understanding of the specific effect of adding or removing multiple edges from a colored graph.
A related question was addressed in Ref.~\cite{Le22} for different quantum sets.

We start in \cref{Cgraphappr} by showing how to  associate with each Bell or NC inequality a (colored) graph and how a (colored) graph can be associated with classical and quantum sets of allowed behaviors. 
Then, in \cref{Cchsh} we apply these graph approaches to the CHSH inequality. 
In \cref{Cresults} we introduce the family of graphs which later on we will show to have quantum sets that are supersets of the colored CHSH graph and  subsets of the simple shadow graph.
Therefore, their quantum sets are not equal and the colored graph approach is indeed a better description of the Bell CHSH inequality.
 We also discuss how slight changes of the graph changes the corresponding quantum sets.
We provide concluding remarks in \cref{Cconclusion}.

\section{Graph Approach} \label{Cgraphappr}

Bell and noncontextuality  inequalities can be represented as graphs. 
Here we will first introduce the CSW (simple) graph approach \cite{CSW1, CSW2} and later the colored-graph approach for bipartite Bell and NC inequalities, as presented in Ref.\ \cite{MultGraphAppr}.
More details, generalizations, and other approaches can be found in Refs.\ \cite{MA,GraphBook,ContexReview}.

In a NC test, measurements of certain contexts are collectively performed. 
A context $\CC = \left\{ x, \ldots, z \right\}$ is a set of jointly compatible measurements $x, \ldots, z$. 
In general these measurements can be performed in a single laboratory or by different parties in separate laboratories. 
In the case of Bell tests, the allowed contexts contain at most one measurement per party. 
The occurrence of measuring a context $\CC = \left\{ x, \ldots, z \right\}$ and getting outcomes $\left\{ a, \ldots, c \right\}$ is called a (measurement) event.
It is denoted by $a, \dots{,} c \vert x, \dots{,} z$.
Two events $v$ and $v'$ are exclusive if both include some measurement $x$ with distinct outcomes $a \neq a'$ \cite{GraphBook}.
Examples of exclusive events are $v = a b \vert x y$ and $v' = a' b' \vert x y$, where the measurements $x$ and $y$ lead to different outcomes $a \neq a'$ and $b \neq b'$, respectively. The event $v'' = a''  b' \vert  x'' y$ is exclusive to $v$ as well, if the measurement $y$ leads to different outcomes $b \neq b'$.

The exclusivity structure of a set of events $\left\{ v_i \right\}_{i=0}^{n-1}$ can be represented by exclusivity graphs.
\begin{definition} \label{dexgraph}
Let $\left\{ v_i \right\}_{i=0}^{n-1}$ be a set of events, describing a NC test. 
Their exclusivity graph is the $n$-vertex graph $G = (V,E)$ where each event corresponds to a vertex $v_i \in V$ and $\{v_i,v_j\} \in E$ whenever the events $v_{i}$ and $v_{j}$ are exclusive. 

A weighted exclusivity graph $(G,\omega)$ is an exclusivity graph which has a weight $\omega_i \geqslant 0$ associated with every vertex $v_i \in V$. 
\end{definition}

\begin{figure}
\centering
\includegraphics[scale=1]{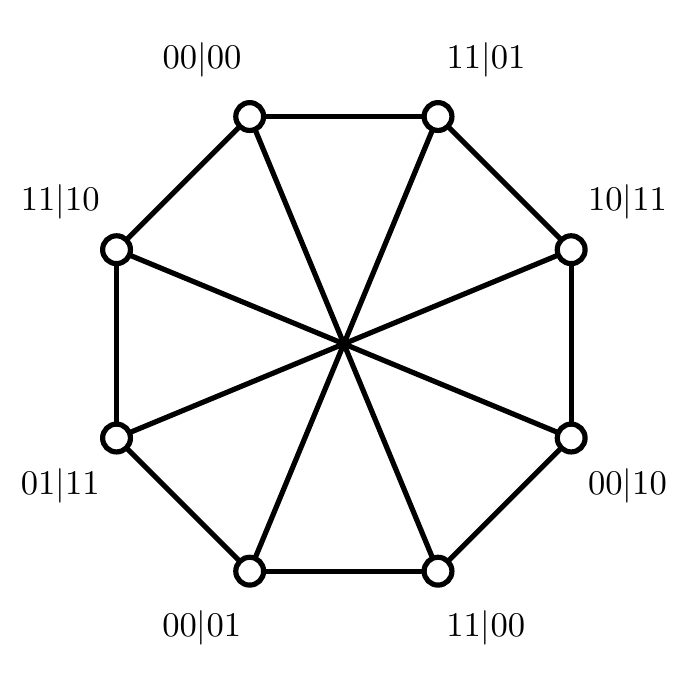}
\caption{\label{Fgsimple}Exclusivity graph $  G = (V, E)$ of the CHSH inequality. 
The events appearing in \cref{EchshP} are represented by vertices and edges represent their exclusivity structure.} 
\end{figure}

The  exclusivity graph representing the CHSH inequality \cite{CHSH} is shown in \cref{Fgsimple}. 
The CHSH inequality is explained in more detail in \cref{Cchsh}.
Other examples of exclusivity graphs can be found in the cited references.

\begin{figure}\label{sfig:EdgeColours}
\centering
\includegraphics[scale=1]{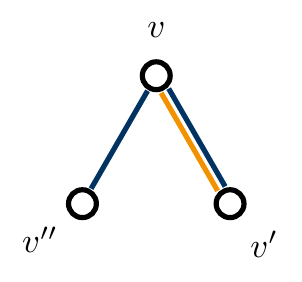}
\caption{\label{Fevent}{Events $v = a b \vert x y$, $v' = a' b' \vert x y$, and $v'' = a''  b' \vert  x'' y$, where $a \neq a'$, $b \neq b'$, and so on, are represented by vertices. Events $v$ and $v'$ are exclusive for both parties. 
This is represented by a double edge [orange (light) and blue (dark)]. 
The events $v$ and $v'$ are exclusive for the second party, only. 
They are therefore connected by a blue (dark) edge.
The events $v'$ and $v''$ are not exclusive and therefore not connected.}} 
\end{figure}

Simple graphs treat every exclusivity  in the same way. 
In some cases it is reasonable to distinguish among different kinds of exclusivity.
In a bipartite scenario, there are three types of exclusivity: exclusivity coming from the first party, from the second party, and from both parties. 
Considering these multiple possibilities, the exclusivity structure of a set of events $\left\{ v_i \right\}_{i=0}^{n-1}$ can be represented by an edge-colored exclusivity graph \cite{MultGraphAppr}.
More precisely, we have the following definition, which is illustrated in  \cref{Fevent}.

\begin{definition}
Let $\left\{ v_i \right\}_{i=0}^{n-1}$ be a set of events, describing a bipartite {NC} test. 
Their bicolored exclusivity graph is the $n$-vertex colored graph $\GG = (V,(E_A,E_B))$ where each event corresponds to a vertex $v_i \in V$ and edge sets $E_A$ and $E_B$ such that $\{v_i,v_j\} \in E_A$ whenever the events $v_{i}$ and $v_{j}$ are exclusive in the first party and $\{v_i,v_j\} \in E_B$ whenever the events $v_{i}$ and $v_{j}$ are exclusive in the second party. 
Edges of different sets are represented by different colors.  

A weighted bicolored exclusivity graph $(\GG,\omega)$ is a bicolored exclusivity graph which has a weight $\omega_i \geqslant 0$ associated with every vertex $v_i \in V$. 

With any (weighted) bicolored graph we can associate a simple (weighted) graph, as in \cref{dexgraph}, with the same vertices (and weights) and edge set $E = E_A \cup E_B$. This simple graph is called the shadow of the colored graph.
\end{definition}

\begin{figure}
\centering
\includegraphics[scale=1]{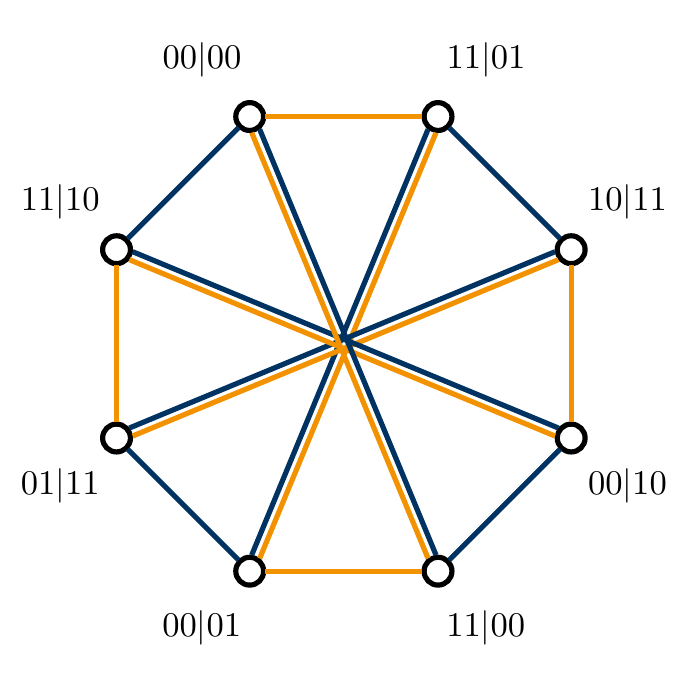}
\caption{\label{Fgchsh}Exclusivity bicolored graph $  \Gchsh = (V, (E_A, E_B))$ of the CHSH inequality. 
The events appearing in \cref{EchshP} are represented by vertices and edges represent their exclusivity structure.
Two parties, Alice and Bob, are represented by two colors: orange (light) and blue (dark), respectively. 
Orange (light) edges represent Alice's exclusivity structure, while blue (dark) edges represent Bob's exclusivity structure.} 
\end{figure}

In this paper we use calligraphic letters for colored graphs and italic letters for simple graphs.
The bicolored exclusivity graph {$\Gchsh$},  representing the CHSH inequality \cite{CHSH}, is shown in \cref{Fgchsh}.
Other examples of colored exclusivity graphs can be found in
 Ref. \cite{MultGraphAppr}. 
 Colors were added to mark the origin of the exclusivity, but one question must be asked: Do the colors in the CHSH graph really imply physically relevant restrictions? 
 In other words, is there anything allowed by the shadow graph of \cref{Fgsimple} which is forbidden by the colored graph of \cref{Fgchsh}?
In this paper we will denote the simple graph of Fig.~\ref{Fgsimple} by  $\Gshad$, since it was introduced by Cabello, Severini, and Winter.
The graph $\Gshad$ represents the NC structure of the CHSH inequality while the colored graph $\Gchsh$ represents the Bell structure of the CHSH inequality.
We will see later, as an important step in this paper, 15 different bicolored exclusivity graphs sharing this same graph $\Gshad$ as their shadows.

A Bell or a NC inequality can be written as a positive linear combination $S$ of probabilities of events and a number $\alpha$ which is the maximum classically 
reachable value within the given inequality
\begin{align}
S(\GG,\omega) = \sum_{i=0}^{n-1} \omega_i P_i \leqslant \alpha (\GG,\omega), \label{Encineq}
\end{align}
where $P_i \coloneqq P(v_i)$ is the probability to obtain event $v_i \in \left\{ v_i \right\}_{i=0}^{n-1}$ and $\omega$ is a weight vector with components $\omega_i \geqslant 0$.
A vector $P \in \Rd^n$ with entries $P_i$ is called the behavior of the test or of the graph. 

As the notation suggests, the classical bound $\alpha(\GG,\omega)$ as well as other properties can be obtained from the graph. 
 The bound $\alpha(\GG,\omega)$ corresponds to the independence number of the graph.
The definition of the independence number is based on the concept of independent sets of a graph. 
 An independent set $I \subseteq V $ of a given graph $\GG= (V,(E_A,E_B))$
 is a set of vertices, which are not adjacent, that is for all pairs of elements $v_i, v_j \in I$ it holds that $(v_i,v_j) \notin E_J$.

\begin{definition}[independence number $\alpha(\GG,\omega)$ \cite{GraphBook}] \label{DindepNr}
The weighted independence number $\alpha(\GG,\omega)$ of a vertex-weighted colored graph $(\GG,\omega)$ is
\begin{align}
\alpha(\GG,\omega) \coloneqq \max_{I \subseteq V} \sum_{i \in I} \omega_i,
\end{align}
where the maximum is taken over all independent sets $I \subseteq V$.
\end{definition}

Independent sets allow us to build the important
 set of classical behaviors $\CC(\GG)$. 
 The set $\CC(\GG)$ is the convex hull of all characteristic vectors of the independent sets (hence a polytope). 
For each independent $I \subseteq V$, its characteristic vector $x^I \in \Rd^n$ has components $x^I_i = 1$ if $v_i\in I$ and $x^I_i = 0$ otherwise.  
The interpretation of $\CC(\GG)$ is the following: Independent sets of events are those deterministic choices allowed by exclusivities;
the only other possibilities are their convex combinations, interpreted as probabilistic mixtures of those allowed configurations.
Adapting and adopting the notation from simple-graph theory, the set $\CC(\GG)$ can also be denoted by $\mathrm{STAB}(\GG)$ \cite{LOVASZ1994137}. 
The weighted independence number $\alpha(\GG,\omega)$ is the maximal value of $S(\GG,\omega)$ attainable 
on $\CC(\GG)$.
When every weight is 1, we use $\alpha(\GG)$ and call it the independence number of $\GG$.
This number is also the cardinality  of the largest independent set of $\GG$, justifying its name.
All this discussion of independent sets, classical polytope, and (weighted) independence number remains the same for simple graphs.
In other words, if $\GG$ is a colored graph and $G$ its shadow, since they originate from the same independent sets, their classical sets also coincide: $\CC(\GG) = \mathrm{C(G)}$ [in graph-theory notation, $\mathrm{STAB}(\GG) = \mathrm{STAB}(G)$].

Now we move on to the relevant definitions for the quantum upper bounds, $\vartheta(G)$ and $\vartheta_c(\GG)$, and the quantum sets $\mathrm{Q(G)}$ and $\QQ(\GG)$.
\begin{definition}[orthogonal labeling for simple graphs] \label{DorthLab}
Let $\left\{ \Pi_i \right\}_{i=0}^{n-1}$  be a set of projectors acting on a finite-dimensional vector space with an inner product.
This set is {an} orthogonal labeling of a simple graph $G = (V,E)$ if each $\Pi_i$ is associated with the vertex $v_i \in V$  and the projectors are orthogonal, that is, $\Pi_i \Pi_j = 0$, whenever
 $\{v_i,v_j\} \in  E$.
\end{definition}
The Lov\'asz number plays a central role in the CSW graph approach to quantum contextuality.
One good definition for this number is the following (see Ref.\ \cite{sandwich} for many others).
\begin{definition}[Lov\'asz number] \label{Lovasz}
Given a {vertex-weighted graph}, $(G, \omega)$, its Lov\'asz number is given by
\begin{align}
\vartheta (G,\omega) \coloneqq \sup \sum_{i = 0}^{n-1} \omega_i \expval{\Pi_i}{\Psi}, \label{Elovasz}
\end{align}
where $\Pi_i$ are projectors from an orthogonal labeling of {the graph} $G$, $\ket{\Psi}$ is a normalized vector, and the supremum is taken over all possible normalized vectors $\ket{\Psi} $ and orthogonal labels $\left\{ \Pi_i \right\}_{i=0}^{n-1}$.
\end{definition}
In graph theory, the vector $\ket{\Psi}$ is called the handle of the representation \cite{Lovasz}. In the following, we will use the term, when we talk about the vector, which maximizes \cref{Elovasz} [as well as \cref{Elovaszc}, in the colored case].
It is not difficult to show that this optimization process can be done using unidimensional projectors, usually also represented by vectors, instead of projectors.

The definition of the Lov\'asz number comes as the maximization from the function $S(G,\omega)$ over a set where $P_i = \expval{\Pi_i}{\Psi}$.
In graph theory, this {set} is the Gr\"otschel-Lov\'asz-Schrijver theta body, denoted by $\Th(G)$.
Since we can naturally identify the handle $\ket{\Psi}$ with a quantum state and the projectors of the representation with effects of dichotomic projective measurements, in quantum theory, the vectors $P = \left(P_i\right)_{i=0}^{n-1}$ obtained in this form are called quantum behaviors.  
The set of all possible quantum behaviors for an exclusivity scenario given by $G$ is called the quantum set of $G$, $\mathrm{Q(G)}$.
One of the most beautiful and important results in the CSW approach to contextuality is this identification: $\mathrm{Q(G)} = \Th(G)$.

\begin{definition}[orthogonal labeling for bicolored graphs] \label{DorthLabMult}
Let $\left\{ \Pi_i = \Pi_i^A \otimes \Pi_i^B \right\}_{i=0}^{n-1}$  be a set of projectors acting on a finite-dimensional vector space with an inner product and a tensor product structure.
This set is an orthogonal labeling of a bicolored graph $\GG = (V,(E_A,E_B))$ if each $\Pi_i$ is associated with a vertex $v_i \in V$  and  the projectors are orthogonal according to the exclusivities, that is, $\Pi_i^J \Pi_j^J = 0$ whenever $\{v_i,v_j\} \in E_J$.
\end{definition}

\begin{definition}[Lov\'asz number for colored graphs] \label{Dlovasz}
Given a vertex-weighted colored graph $(\GG, \omega)$, its (colored) Lov\'asz number is given by
\begin{align}
\vartheta_c (\GG,\omega) \coloneqq \sup \sum_{i = 0}^{n-1} \omega_i \expval{\Pi_i}{\Psi}, \label{Elovaszc}
\end{align}
where $\Pi_i = \Pi_i^A \otimes \Pi_i^B$ are  projectors from an orthogonal labeling, $\ket{\Psi}$ is a normalized vector, and the supremum is taken over all possible normalized vectors $\ket{\Psi}$ and orthogonal labeling $\left\{ \Pi_i \right\}_{i=0}^{n-1}$.
\end{definition}

In both cases, if every weight equals $1$, we get the corresponding (color) Lov\'asz number of the respective graphs, $\vartheta(G)$ and $\vartheta_c(\GG)$.

The colored Lov\'asz number $\vartheta_c$ is the best possible upper bound to the quantum bound of the underlying Bell or NC inequality with parts \cite{MultGraphAppr,slofstra_2019}. 
The simple Lov\'asz number, $\vartheta$, of its shadow graph is a not necessarily tight upper bound for the same quantity \cite{CSW2, GraphBook}.

As in the previous case, the weighted colored Lov\'asz number comes as the maximization of $S(\GG,\omega)$ at a set of vectors $P = (P_i)_{i=0}^{n-1}$, where $P_i = \expval{\Pi_i}{\Psi}$ now with one extra restriction: $\Pi_i = \Pi_i^A \otimes \Pi_i^B$.
It corresponds to the set of quantum behaviors obeying the bipartite restrictions of the colored graph $\GG$.
For this reason, it is identified with the quantum set of the colored graph $\QQ(\GG)$. 
{Following the language of graph theory, we could also call this set the colored theta body $\Tg$. 
If we were only concerned with one $P_i$, since there is no restriction on dimensions, we could include the image of the uncolored $\Pi_i$ as a subspace of a colored one and obtain the same $P_i$ using a handle orthogonal to the complement of such subspace.
It is not clear, however, whether or when  it is possible to use such a trick simultaneously for every $P_i$ of a given graph. }

A final comment is in order.
Since we are only considering pure states and projective-valued measurements (PVMs), this definition might look too restrictive to be  associated with the quantum set. 
Since we are dealing only with two-outcome measurements, Appendix B of Ref.~\cite{Irfan20} can be used to simultaneously dilate compatible measurements including PVMs, in the sense of Naimark dilation theorem \cite{peres1990neumark, NielsenChuang}.
State purification can also be used, presenting a higher-dimensional system where PVMs and pure states emulate the same behavior. A detailed discussion of this can be found in \cref{Sapp_dilation}.

\section{The CHSH Bell-graph and its NC-shadow-graph} \label{Cchsh}

The most popular Bell inequality is the CHSH inequality \cite{CHSH}.
In many textbooks it is represented as 
   \begin{align}
	 \tilde S_{\text{CHSH}} = \left\langle
	x_0 y_0  + x_0 y_1  +
	x_1 y_0  - x_1 y_1 \right\rangle  \leqslant 2, \label{Echsh}
   \end{align}
where $x_0$, $x_1$, $y_0$, and $y_1$ are dichotomic random variables which can take values $\pm 1$ at each run. 
These variables can be seen as measurement results of two parties, where one party, Alice, chooses from a measurement set $X = \lbrace x_0, x_1 \rbrace $ and the second party, Bob, chooses from a measurement set $Y = \lbrace y_0, y_1 \rbrace$.
In classical theories, the maximal value which can be reached is 2. Using quantum measurements $x$ and $y$ on distinct systems, the expectation value of measuring a state $\ket{\Psi}$ is given by  {$\expval{x \otimes y}{\Psi}$}. All involved measurements are two-outcome measurements.
The maximal quantum value goes beyond the classical bound  and is given by the Tsirelson bound $2\sq$ \cite{Tsirelson}.
This inequality can also be written in the form \eqref{Encineq} as
\begin{align}
\Schsh &= \mathmakebox[\widthof{+}]{} P(00\vert 00) + P(00\vert 01) + P(00\vert 10)   \notag \\ 
& \mathmakebox[\widthof{=}]{}+ P(01\vert 11) + P(11\vert 00) + P(11\vert 01) \notag \\  
& \mathmakebox[\widthof{=}]{} + P(11\vert 10) + P(10\vert 11) \leqslant 3, \label{EchshP} 
\end{align}
where $P(v)$ is the probability of obtaining the event $v$.
In this form, the inequality can be represented as an exclusivity graph.
The colored exclusivity graph of the inequality in \eqref{EchshP} is shown in \cref{Fgchsh}. 
We refer to this graph as CHSH Bell graph $\Gchsh$, while the simple graph $\Gshad$ of \cref{Fgsimple} representing only the NC exclusivities will be referred as the CHSH NC graph.
The independence number of  both graphs is $\alpha = 3$ and the simple as well as the colored Lov\'asz number is $\vartheta = \vartheta_c = 2 + \sq$, which
corresponds to the classical and quantum maximal values of the CHSH inequality in  \eqref{EchshP}, respectively.

In general, a colored graph better describes the underlying test than the simple graph. 
This results in a colored Lov\'asz number which is smaller than or equal to the Lov\'asz number of its shadow graph and therefore can give a tighter bound to the maximal quantum value. 
In the case of the CHSH inequality, the  NC graph $\Gshad$ already gives the precise quantum bound of the inequality.  
In graph terms, the original Lov\'asz number coincides with its colored version: $\vartheta(\Gshad) = \vartheta_c(\Gchsh)$.

The coincidence between the Lov\'asz number of the  CHSH NC graph $\Gshad$ and its colored version $\Gchsh$ says that the maximal quantum value coincides whether we see the CHSH inequality as a Bell or a NC inequality. 
Is it true that any quantum correlation allowed by $\Gshad$ can also be obtained in the more restrictive colored version $\Gchsh$?
We show  not only that the quantum set of the NC graph $\Qshad$ allows for a larger quantum set than the quantum set of the colored CHSH Bell graph $\Qchsh$, but also that there are many other colored graphs with the same shadow which generate intermediate quantum sets. 
We are interested in exploring the differences among those sets.

\section{Results} \label{Cresults}

\begin{figure*}%[h]
\includegraphics[scale=1]{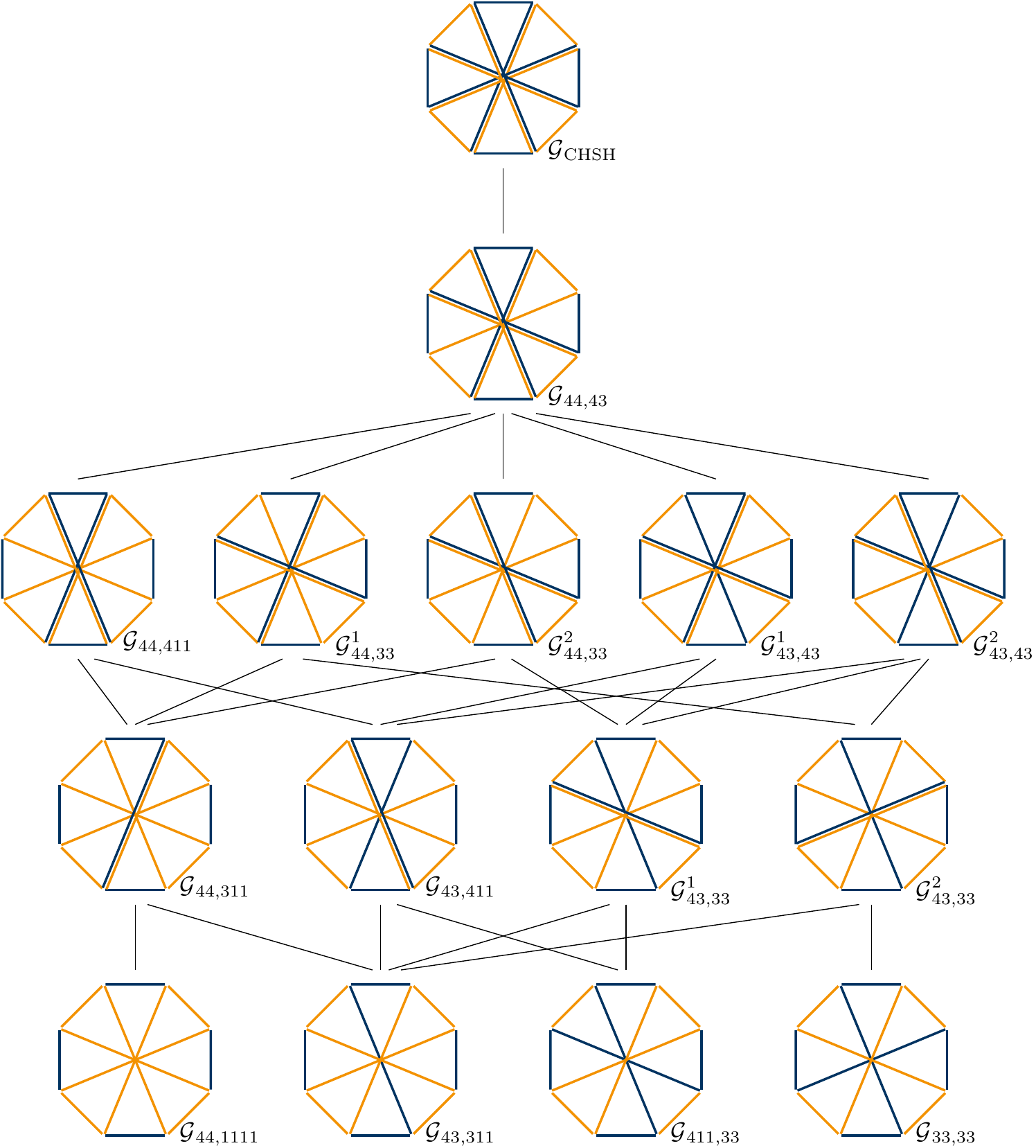} 
\caption{Family of 15 different colored graphs with the same shadow $\Gshad$. 
On the top is the graph $\Gchsh$ with four double edges. 
One row below is the graph $\GG_{44,43}$, which can be obtained from $\Gchsh$ by removing one edge (and using a color-graph isomorphism, if necessary).
The next level shows the five different possibilities of removing an edge from $\GG_{44,43}$, preserving the shadow $\Gshad$. 
The following level lists all the graphs obtained in this process which have just one double edge, while the bottom level shows the four subgraphs of $\Gchsh$ with shadow $\Gshad$ and no double edges. 
There is a line between a graph in one level and another graph in the row below if the latter can be obtained by removing one edge from the former (and possibly a change of colors).}
 \label{Fall_graphs}

\end{figure*}

We first state some general properties of any quantum set.
The quantum set is convex. 
Moreover, for a behavior $P \in \QQ$, every behavior $P' \in \Rd^{\lvert V \rvert}$ where $P'_i \leqslant P_i$ for all $v_i \in V$ is fulfilled is a behavior of $\QQ$ as well  \cite{concavity}.
  We are therefore only interested in finding behaviors on the boundary of the quantum set. Other useful properties are given in \cref{Trotation,TRrotation}.

\begin{remark} \label{Trotation}
The quantum set $\QQ$ of a weighted graph $\left(\GG, \omega \right)$ is independent of the weight $\omega$.
\end{remark}

\begin{theorem} \label{TRrotation}
Let $\vartheta_c \left(\GG, \omega \right)$ be the colored Lov\'asz number of a weighted graph $\left(\GG, \omega \right)$. If there exist an orthogonal labeling $\left\{ \Pi_i \right\}$ of $\left(\GG, \omega \right)$ and a handle $\ket{\Psi}$, such that $\sum_{i = 0}^{n-1} \omega_i \expval{\Pi_i}{\Psi} = \vartheta_c \left(\GG, \omega \right)$, 
the behavior $P$ induced by this orthogonal labeling and handle is on the boundary of the quantum set $\QQ \left(\GG \right)$.
\end{theorem}

\cref{Trotation} comes from the fact that only exclusivities play a role in the definition of quantum behaviors. 
\cref{TRrotation} follows from the linearity of the function being optimized and the convexity of $\QQ(\GG)$.
Since every weighted graph induces an inequality, the family of weighted graphs  $(\GG,\omega)$ sharing the same quantum set $\QQ(\GG)$ induces a family of inequalities $S(\GG,\omega) = \sum_{i = 0}^{n-1} \omega_i P_i \leqslant \vartheta_c (\GG,\omega)$, actually defining the quantum set of $\GG$.
 It is noteworthy that these inequalities are maximized by in general different quantum behaviors, since $\omega$ defines a direction in the space where behaviors are defined. 
 By finding behaviors which maximize $S(\GG,\omega)$, we can find  behaviors on the boundary of $\QQ(\GG)$.

We are interested in how and why the quantum set of the graph $\Gchsh$ differs from the quantum set of its shadow graph $\Gshad$.
In order to compare the CHSH Bell quantum set $\Qchsh$ with the CHSH NC  quantum set $\Qshad$, we introduce a family of colored graphs whose quantum sets are supersets of $\Qchsh$ but subsets of the set $\Qshad$. 

It is technically easier to compare colored graphs with colored graphs. 
We will show that there are colored graphs whose quantum sets are subsets of the quantum set of the CHSH NC graph $\mathrm{Q}(\Gshad)$ but supersets of the quantum set of the CHSH Bell graph $\QQ(\Gshad)$. It also gives us hints how specific edges influence the behavior of a graph.
This observation, which is essential in our approach, is summarized in the following remark.
\begin{remark} \label{Rsubgraphs}
Let $\GG' = (V, E')$ and $\GG = (V, E)$ be  two graphs with $E' \subseteq E$ such that $\GG'$ is a subgraph of $\GG$. Then the quantum set $\QQ(\GG)$ is a subset of the quantum set of $\GG'$:
$ \QQ(\GG) \subseteq \QQ(\GG')$.
\end{remark}
\cref{Rsubgraphs} comes from the fact that edges of the graph are restrictions on its quantum set. 
Removing edges from a graph is therefore equivalent to having fewer restrictions on the quantum set. 
This usually allows for a larger quantum set.
Note additionally  that the quantum set of a colored graph always lays inside the quantum set of its shadow graph.

\cref{Fall_graphs} shows the family of subgraphs of $\Gchsh$ which all have the same shadow. 
There are 15 graphs which are different among each other up to colored graph isomorphisms. 
We introduce a notation to distinguish among them: 
For each color we count the number of edges in each component of the graph and write them as indices. We use commas to separate between colors. 
For example the graph $\GG_{44,311}$ denotes a bicolored graph where the graph of the first color contains two non-adjacent subgraphs with four edges each and the second color graph contains three non-adjacent subgraphs with 3, 1, and 1 edges, respectively.
In the cases where this notation is not sufficient to select just one graph in the family,  we use superscripts to discriminate between them.
Note that the introduced notation is suitable for this family of interest and for the purposes of this paper, but it is not a way of well characterizing colored graphs in general.

In order to compare the sets $\Qchsh$ and $\Qshad$, we use as intermediate sets the quantum sets of some graphs of this family.
We choose the graphs $\Gdrei$ and $\Gvier$, since they are both very distant from $\Gchsh$ and from each other in the genealogic tree of \cref{Fall_graphs}. 
Using the Navascu\'es-Pironio-Ac\'in (NPA) hierarchy \cite{NPA07, NPA08}, we numerically find, for some choices of weights, upper bounds for $\vartheta_c \left(\GG,\omega \right)$. 
These numerical results give upper bounds to the real maximal quantum bound. 
We then approximate points in the boundary by obtaining
 explicit behaviors from orthogonal labelings and handles.  
As it will be shown, for some choices of graph and weight, the agreement of these two approaches is really good. This shows that we know to a pretty good approximation points in the boundary of the corresponding quantum set direction.
 For other cases a small gap is still present, demanding more research.

\begin{figure}
\includegraphics[scale=1]{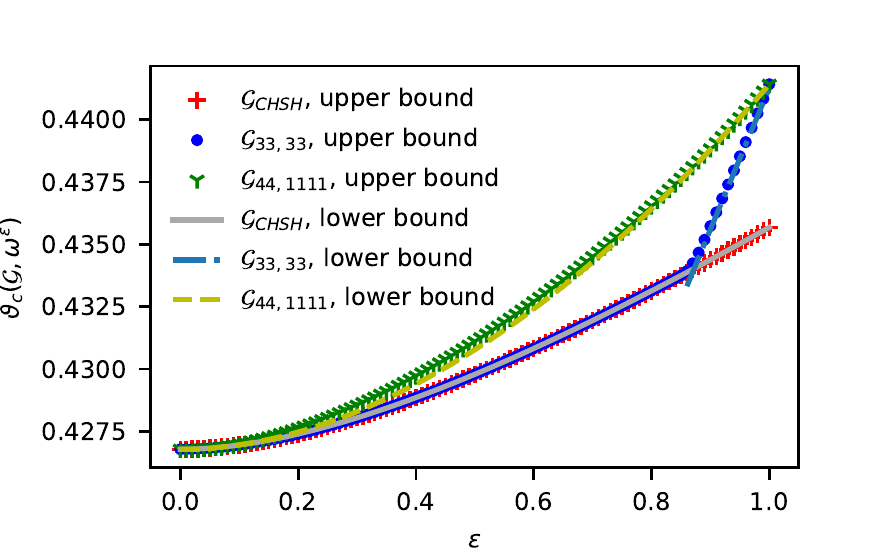}
\caption{Upper and lower bounds of the colored Lov\'asz numbers $\vartheta_c\left(\Gchsh, \omega_5^{\varepsilon}(0^-1^-7^-) \right)$, $\vartheta_c\left(\Gdrei, \omega_5^{\varepsilon}(0^-1^-7^-) \right)$, and $\vartheta_c\left(\Gvier, \omega_5^{\varepsilon}(0^-1^-7^-) \right)$. For $\varepsilon = 0$ we have in all cases $\vartheta_c \approx 0.427$. For $\varepsilon = 1$ we have $\vartheta_{c\ \text{CHSH}} \approx 0.436$ and $\vartheta_{c\ 33,33} = \vartheta_{c\ 44,1111} \approx 0.442$. The curve of $\vartheta_c\left(\Gdrei, \omega_5^{\varepsilon}(0^-1^-7^-) \right)$ has a kink at $\varepsilon \approx 0.85$.} \label{FG_5_compare}
\end{figure}

\cref{FG_5_compare} shows upper and lower bounds of the graphs $\Gchsh$, $\Gdrei$, and $\Gvier$, weighted with the weight vector
  \begin{align}
\omega_5^{\varepsilon}(0^-1^-7^-) 
= (1-\varepsilon)
    \begin{pmatrix}
           \nicefrac{1}{8} \\
           \nicefrac{1}{8} \\
           \nicefrac{1}{8} \\
           \nicefrac{1}{8} \\
           \nicefrac{1}{8} \\
           \nicefrac{1}{8} \\
           \nicefrac{1}{8} \\
           \nicefrac{1}{8} 
     \end{pmatrix}
+ \varepsilon
    \begin{pmatrix}
           0 \\
           0 \\
           \nicefrac{1}{5} \\
           \nicefrac{1}{5} \\
           \nicefrac{1}{5} \\
           \nicefrac{1}{5} \\
           \nicefrac{1}{5} \\
           0 
     \end{pmatrix}
,
 \label{Eomega_5}
  \end{align}
where $\varepsilon \in \left[ 0, 1 \right]$. {Note, that applying $\omega_5^{1}(0^-1^-7^-)$ to the given graphs and comparing graphs of the pentagonal inequalities, given in Ref.~\cite{MultGraphAppr}, we can directly answer our first question: There are behaviors in $\Qshad$ which are not in $\Qchsh$. We want to understand better, where the quantum boundaries are equal and where they differ.}

It is noteworthy that we found the given curves of $\Gchsh$ and $\Gvier$ with orthogonal labeling $\left\{ \Pi_i \right\}$ and a handle $\ket{\Psi}$ in Hilbert spaces $\HH^2 \otimes \HH^2$, while we need a Hilbert space $\HH^2 \otimes \HH^3$ in order to approximate the curve of $\Gdrei$.
This is consistent with the fact that the first and second graphs give Bell inequalities which can be maximally violated by two qubits, while the former only gives a generalized Bell inequality, demanding extra dimensions in one part \cite{building_blocks}.
In the case $\varepsilon = 0$, all graphs have the same Lov\'asz number which can be reached with the orthogonal labeling and handle known from the case of CHSH inequality: $\left\{ \Pi_i^{\text{CHSH}} \right\}$ and $\ket{\Psi^{\text{CHSH}}}$. For $\varepsilon = 1$, the graphs reduce to the graphs of the first and third pentagonal inequalities $I_1^P$ and $I_3^P$, respectively, where the optimal solutions are given in \cite{PentBI}. {Note that we introduced the labels $I_1^P$ and $I_3^P$ to refer to the inequalities in Ref.\ \cite{PentBI}, which were originally called first Bell inequality and third Bell inequality, respectively.} For $\Gchsh$ and $\Gvier$, we approximate the analytic curve by a superposition of $\ket{\Psi^{\text{CHSH}}}$ and $\ket{\Psi^{I_{1}^P}}$ or $\ket{\Psi^{I_{3}^P}}$ and rotations of the projectors from $\Pi_i^{\text{CHSH}}$ to $\Pi_i^{I_{1}^P}$ and $\Pi_i^{I_{3}^P}$, respectively. 
In the case of $\Gdrei$, we can approximate the curve with a superposition of $\ket{\Psi^{\text{CHSH}}}$ and $\ket{\Psi^{I_{3}^P}}$ and rotations of the projectors from $\Pi_i^{\text{CHSH}}$ to $\Pi_i^{I_{3}^P}$ for $i \neq 1$ and $\Pi_1 = \dyad{22}$.
A more detailed phenomenological description of the results can be found in \cite{MA} and in \cref{Sapp_detail}.
Using the same method we can analyze other paths
with the goal to generalize to manifolds on the boundary which could be described parametrically.

\begin{figure*}%[b]
\centering
\includegraphics[scale=1]{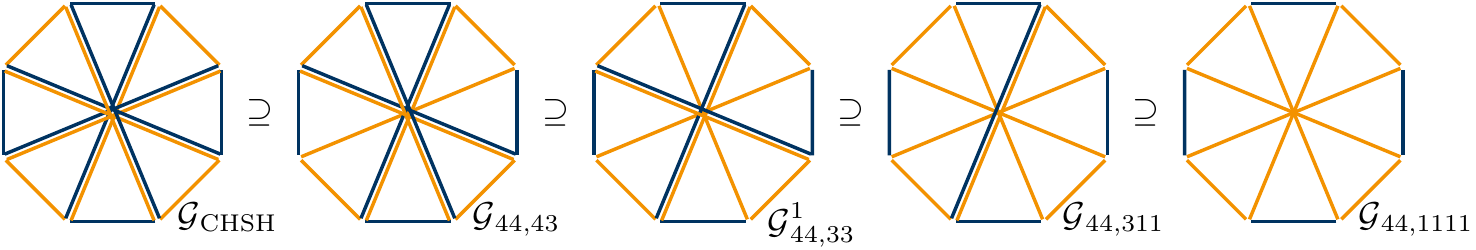}
\caption{Chain of subgraphs. The chain is constructed as follows: We start with $\Gchsh$ (on the left) and remove edges one at a time until we reach $\Gvier$ (on the right). Each graph is a subgraph of all graphs on its left side. Equivalently, each quantum set of a graph is a subset of all quantum sets of graphs on its right side.} \label{Fchain_of_graphs}
\end{figure*}

A related interesting question is how changes in the graph  influence the quantum set. 
To explore these changes, we compare the upper bound of colored Lov\'asz numbers of the chain of graphs given in \cref{Fchain_of_graphs} for certain weight vectors $\omega$.

\begin{figure}
\includegraphics[scale=1]{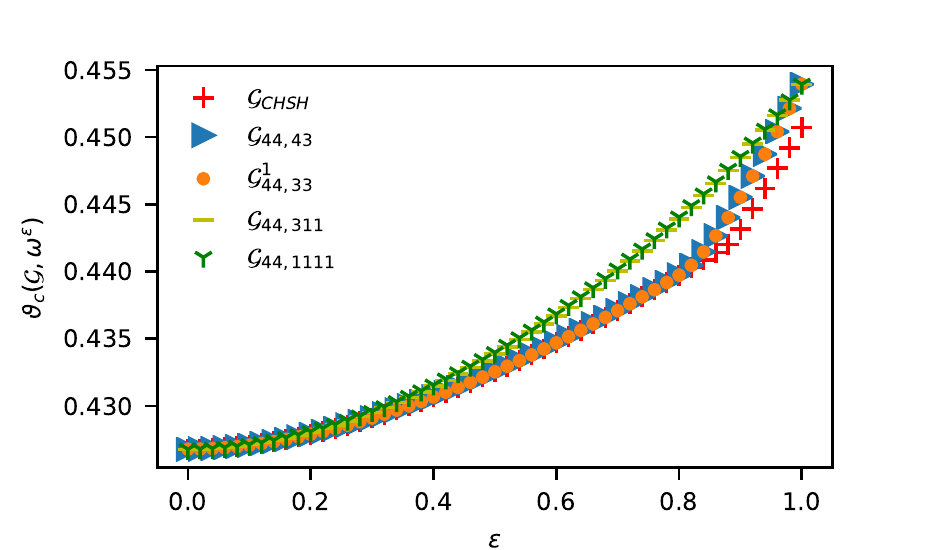}
\caption{Upper bounds of the colored Lov\'asz numbers of the graphs given in \cref{Fchain_of_graphs}, weighted with $\omega^{\varepsilon}$ as given in \cref{Eomega_5_rand}. For $\varepsilon = 0$ we have in all cases $\vartheta_c \approx 0.427$. Four of the five curves are different and therefore the quantum sets of the underlying graphs are different, as well. Some explicit numbers are given in \cref{TG_5_chain}.} \label{FG_5_chain}
\end{figure}

In order to understand how removing one edge in the graph changes the quantum set,
we start proceeding in the same way as before, and surprisingly the five graphs of this chain only generate three distinct curves.
Since symmetry is closely related to degenerescence, we then move to the weight vectors
  \begin{align}
\omega_5^{\varepsilon}(0^-1^-7^-) 
= (1-\varepsilon)
    \begin{pmatrix}
           \nicefrac{1}{8} \\
           \nicefrac{1}{8} \\
           \nicefrac{1}{8} \\
           \nicefrac{1}{8} \\
           \nicefrac{1}{8} \\
           \nicefrac{1}{8} \\
           \nicefrac{1}{8} \\
           \nicefrac{1}{8} 
     \end{pmatrix}
+ \varepsilon
    \begin{pmatrix}
           0 \\
           0 \\
           \kappa_2 \\
           \kappa_3 \\
           \kappa_4 \\
           \kappa_5 \\
           \kappa_6 \\
           0 
     \end{pmatrix}
,
 \label{Eomega_5_rand}
  \end{align}
with random variables $\kappa_i \geqslant 0$, such that $\sum_{i=2}^6 \kappa_i = 1$ and $\varepsilon \in \left[ 0, 1 \right]$. One example with random variables $( \kappa_i )_{i=2}^6 \approx ( 0.26, 0.18, 0.19, 0.13, 0.24 )$ is shown in \cref{FG_5_chain}. 
Some explicit results of the colored Lov\'asz numbers $\vartheta_c (\GG, \omega^{\varepsilon})$ of this example are given in \cref{TG_5_chain}. We can see, that for most graphs, there are some $\varepsilon$ such that the colored Lov\'asz number is greater than the colored Lov\'asz number of its supergraphs and therefore the quantum sets are indeed different. 
The colored Lov\'asz numbers of $\GG_{44,43}$ and $\GG_{44,33}^1$ appear to be the same for the given choice of $( \kappa_i )$ and every value of $\varepsilon$ we tested. 
This indicates that the boundaries of the quantum sets of these two colored graphs share this manifold. 
However, these two quantum sets do not coincide. A different choice of weights can distinguish them, as shown in Ref.\ \cite{supplement},
where examples with different random numbers can also be found.

\begin{table}%[h]
\centering
\begin{tabular}{  c | c  c  c c c} 
$\varepsilon$  & $\Gchsh$ &  $\GG_{44,43}$ & $\GG_{44,33}^1$ & $\GG_{44,311}$   & $\Gvier$   \\ \hline\hline
 
0.3 & 0.4292 & 0.4292 & 0.4292 & 0.4296 & 0.4296 \\ 
0.5 & 0.4326 & 0.4326 & 0.4326 & 0.4339 & 0.4340 \\ 
0.9 & 0.4432 & 0.4456 & 0.4456 & 0.4485 & 0.4486 \\  
\end{tabular}
\caption{Some explicit numbers $\vartheta_c (\GG, \omega^{\varepsilon})$ of the curves in \cref{FG_5_chain}, rounded to the fourth digit.
} \label{TG_5_chain}
\end{table}

We make the same calculations for several more weight vectors and see that in all cases the colored Lov\'asz numbers $\vartheta_c (\Gdrei, \omega)$ are smaller than or equal to $\vartheta_c (\Gvier, \omega)$. 
This may indicate that the quantum set $\Qdrei$ is a subset of the set $\Qvier$. 
This is surprising since the structures of the graphs $\Gdrei$ and $\Gvier$ are very different and $\Gdrei$ is not subgraph of $\Gvier$. 
We uploaded all plots we made, as well as the 
 code we used for the simulations, in Ref.\ \cite{supplement}.

\section{Outlook and Conclusions} \label{Cconclusion}

In this work we reviewed the simple and the multicolored graph approaches to non-contextuality inequalities  \cite{MultGraphAppr} and focused on one of its open questions: 
Is the theta body of the simple CHSH NC graph $\Tshad$ equal to the quantum set of the CHSH Bell scenario $\Qchsh$?
We further explored the boundaries of quantum sets $\QQ(\GG)$ by analyzing various linear functions. We used that when
 two different colored graphs $\GG$ and $\GG'$, where $\GG'$ was obtained by  removing some edges from $\GG$, happen to have different colored Lov\'asz numbers for the same weight, $\vartheta_c(\GG,\omega) < \vartheta_c(\GG',\omega)$, this implies that $\QQ(\GG) \subsetneq \QQ(\GG')$.
By varying weights according to a parameter, we could essentially move on the boundaries of different quantum sets and plot the graphs that show many cases of proper inclusions.

Our goal was to compare  the quantum set $\Qshad$  of the noncolored CHSH NC graph $\Gshad$ with 
the quantum set $\Qchsh$ of the colored CHSH Bell graph $\Gchsh$ in order to find out whether $\Qshad$ is equal to the quantum set of CHSH $\Qchsh$ and what these sets look like.
Since we show the difference of these sets, we now know that it is possible to propose experiments where  quantum behaviors which can be associated with the shadow CHSH NC graph can be produced, which would be unattainable under the CHSH Bell constraints.

As an intermediate step, we found $14$ colored graphs with intermediate quantum sets $\QQ$, that is $\Qchsh \subset \QQ \subset \Qshad$.
This means that some of the behaviors mentioned above can be obtained in some two-player realization, but again, not under CHSH Bell constraint of two parties, two possible measurements for each part, and two possible outcomes for each measurement.
Two such sets are $\Qdrei$ and $\Qvier$. We compared $\Qchsh$ with $\Qdrei$ and $\Qvier$ in the following way:
We first calculated numerically an upper bound to the colored Lov\'asz number $\vartheta_c(\GG, \omega(\varepsilon))$ of the graphs $\Gchsh$, $\Gdrei$, and $\Gvier$ using the NPA hierarchy \cite{NPA07, NPA08}. Then, we computed a lower bound by constructing behaviors $P \in \QQ(\GG)$. 
We showed that the quantum sets of the three graphs are not the same and especially that $\Qdrei$ and $\Qvier$ both are strictly larger than $\Qchsh$. 
The examples which we computed may further indicate that, quite surprisingly, $\Qdrei \subsetneq \Qvier$. 
It would be interesting to verify this observation by a proof or falsify it by finding a weight, $\omega^*$, such that $\vartheta_c(\Qdrei,\omega^*) > \vartheta_c(\Qvier,\omega^*)$.

In order to understand how introducing colors influences the quantum set, we are also interested in how the quantum sets of the different graphs from the family of graphs with the same shadow look.
It is  interesting to know how displacing or removing single edges changes the quantum sets. 
In this paper we computed good upper bounds to colored Lov\'asz numbers of a chain of graphs, where we removed edges one by one, distinguishing among the majority of them, but not all.
Ref.\ \cite{supplement} provides some more details, including other members of the larger family. 
It would be interesting to analyze more paths as well as higher-dimensional submanifolds on the boundary in order to understand the whole structure of the quantum set for each graph.
A natural question still to be answered is to compare the quantum sets of two graphs 
where our notation needs an extra label to distinguish between them, like $\GG_{44,33}^1$ and $\GG_{44,33}^2$.

We are also interested in the Hilbert space dimension of the projectors of the orthogonal labelings and handles which are needed to describe the quantum set. 
In the given examples, it was sufficient to use  projectors and handles in $\HH^2 \otimes \HH^2$ in order to find behaviors of $\Gchsh$ and $\Gvier$. 
For $\Gdrei$ projectors on a Hilbert space $\HH^3 \otimes \HH^2$ are needed \cite{building_blocks}. 
One interesting question remains open: how the colored graph structure indicates when finite dimensions are enough.
At first glance, one could imagine that finite numbers of vertices and colors would imply the possibility of maximizing using finite dimensions, as it happens for CSW graphs. 
However, the inequality $I_{3322}$   is a very good example where a finite graph admits a larger lower bound for the colored Lov\'asz number, with a higher dimension of the space used for the orthogonal representation \cite{Pal10}.
A method to find an upper bound of the needed dimension for a monochromatic graph approach was given in Ref.\ \cite{dimension_wittnessing}. 
This approach may be extended to some multi-colored graphs and used to answer question.

\begin{acknowledgments}
We thank Fabian Bernards, Rafael Rabelo, Elie Wolf, Jan L.~B\"onsel, and Fabian Zickgraf for interesting discussions. We also thank the referees for constructive criticism. To  formulate  and solve the SDPs,  we made use of the package NCPOL2SDPA  \cite{ncpol2sdpa}  and the MOSEK  \cite{mosek}  solver. 
%In order to compute the parameters in \cref{Sapp_detail}, we used Mathematica \cite{mathematica}.
L.V.\ was supported by the Deutsche Forschungsgemeinschaft (DFG, German
Research Foundation, project numbers 447948357 and
440958198), the Sino-German Center for Research
Promotion (Project M-0294), the ERC (Consolidator
Grant 683107/TempoQ), and the Stiftung der Deutschen Wirtschaft. 
M.T.C.\ was partially supported by CNPq Grant No. 310269/2019-9.
This work was supported in part by the Brazilian National Institute of Science and Technology on Quantum Information.

\end{acknowledgments}

\bibliographystyle{apsrev4-2}
\bibliography{verzeichnis3}

%apsrev4-2.bst 2019-01-14 (MD) hand-edited version of apsrev4-1.bst
%Control: key (0)
%Control: author (72) initials jnrlst
%Control: editor formatted (1) identically to author
%Control: production of article title (-1) disabled
%Control: page (0) single
%Control: year (1) truncated
%Control: production of eprint (0) enabled
\begin{thebibliography}{30}%
\makeatletter
\providecommand \@ifxundefined [1]{%
 \@ifx{#1\undefined}
}%
\providecommand \@ifnum [1]{%
 \ifnum #1\expandafter \@firstoftwo
 \else \expandafter \@secondoftwo
 \fi
}%
\providecommand \@ifx [1]{%
 \ifx #1\expandafter \@firstoftwo
 \else \expandafter \@secondoftwo
 \fi
}%
\providecommand \natexlab [1]{#1}%
\providecommand \enquote  [1]{``#1''}%
\providecommand \bibnamefont  [1]{#1}%
\providecommand \bibfnamefont [1]{#1}%
\providecommand \citenamefont [1]{#1}%
\providecommand \href@noop [0]{\@secondoftwo}%
\providecommand \href [0]{\begingroup \@sanitize@url \@href}%
\providecommand \@href[1]{\@@startlink{#1}\@@href}%
\providecommand \@@href[1]{\endgroup#1\@@endlink}%
\providecommand \@sanitize@url [0]{\catcode `\\12\catcode `\$12\catcode
  `\&12\catcode `\#12\catcode `\^12\catcode `\_12\catcode `\%12\relax}%
\providecommand \@@startlink[1]{}%
\providecommand \@@endlink[0]{}%
\providecommand \url  [0]{\begingroup\@sanitize@url \@url }%
\providecommand \@url [1]{\endgroup\@href {#1}{\urlprefix }}%
\providecommand \urlprefix  [0]{URL }%
\providecommand \Eprint [0]{\href }%
\providecommand \doibase [0]{https://doi.org/}%
\providecommand \selectlanguage [0]{\@gobble}%
\providecommand \bibinfo  [0]{\@secondoftwo}%
\providecommand \bibfield  [0]{\@secondoftwo}%
\providecommand \translation [1]{[#1]}%
\providecommand \BibitemOpen [0]{}%
\providecommand \bibitemStop [0]{}%
\providecommand \bibitemNoStop [0]{.\EOS\space}%
\providecommand \EOS [0]{\spacefactor3000\relax}%
\providecommand \BibitemShut  [1]{\csname bibitem#1\endcsname}%
\let\auto@bib@innerbib\@empty
%</preamble>
\bibitem [{\citenamefont {Bell}(1966)}]{Bell66}%
  \BibitemOpen
  \bibfield  {author} {\bibinfo {author} {\bibfnamefont {J.~S.}\ \bibnamefont
  {Bell}},\ }\href {https://doi.org/10.1103/RevModPhys.38.447} {\bibfield
  {journal} {\bibinfo  {journal} {Rev. Mod. Phys.}\ }\textbf {\bibinfo {volume}
  {38}},\ \bibinfo {pages} {447} (\bibinfo {year} {1966})}\BibitemShut
  {NoStop}%
\bibitem [{\citenamefont {Bell}(1964)}]{Bell}%
  \BibitemOpen
  \bibfield  {author} {\bibinfo {author} {\bibfnamefont {J.~S.}\ \bibnamefont
  {Bell}},\ }\href {https://doi.org/10.1103/PhysicsPhysiqueFizika.1.195}
  {\bibfield  {journal} {\bibinfo  {journal} {Physics Physique Fizika}\
  }\textbf {\bibinfo {volume} {1}},\ \bibinfo {pages} {195} (\bibinfo {year}
  {1964})}\BibitemShut {NoStop}%
\bibitem [{\citenamefont {Clauser}\ \emph {et~al.}(1969)\citenamefont
  {Clauser}, \citenamefont {Horne}, \citenamefont {Shimony},\ and\
  \citenamefont {Holt}}]{CHSH}%
  \BibitemOpen
  \bibfield  {author} {\bibinfo {author} {\bibfnamefont {J.~F.}\ \bibnamefont
  {Clauser}}, \bibinfo {author} {\bibfnamefont {M.~A.}\ \bibnamefont {Horne}},
  \bibinfo {author} {\bibfnamefont {A.}~\bibnamefont {Shimony}},\ and\ \bibinfo
  {author} {\bibfnamefont {R.~A.}\ \bibnamefont {Holt}},\ }\href
  {https://doi.org/10.1103/PhysRevLett.23.880} {\bibfield  {journal} {\bibinfo
  {journal} {Phys. Rev. Lett.}\ }\textbf {\bibinfo {volume} {23}},\ \bibinfo
  {pages} {880} (\bibinfo {year} {1969})}\BibitemShut {NoStop}%
\bibitem [{\citenamefont {Klyachko}\ \emph {et~al.}(2008)\citenamefont
  {Klyachko}, \citenamefont {Can}, \citenamefont
  {Binicio\ifmmode~\breve{g}\else \u{g}\fi{}lu},\ and\ \citenamefont
  {Shumovsky}}]{KCBS}%
  \BibitemOpen
  \bibfield  {author} {\bibinfo {author} {\bibfnamefont {A.~A.}\ \bibnamefont
  {Klyachko}}, \bibinfo {author} {\bibfnamefont {M.~A.}\ \bibnamefont {Can}},
  \bibinfo {author} {\bibfnamefont {S.}~\bibnamefont
  {Binicio\ifmmode~\breve{g}\else \u{g}\fi{}lu}},\ and\ \bibinfo {author}
  {\bibfnamefont {A.~S.}\ \bibnamefont {Shumovsky}},\ }\href
  {https://doi.org/10.1103/PhysRevLett.101.020403} {\bibfield  {journal}
  {\bibinfo  {journal} {Phys. Rev. Lett.}\ }\textbf {\bibinfo {volume} {101}},\
  \bibinfo {pages} {020403} (\bibinfo {year} {2008})}\BibitemShut {NoStop}%
\bibitem [{\citenamefont {Sadiq}\ \emph {et~al.}(2013)\citenamefont {Sadiq},
  \citenamefont {Badzi{\k a}g}, \citenamefont {Bourennane},\ and\ \citenamefont
  {Cabello}}]{PentBI}%
  \BibitemOpen
  \bibfield  {author} {\bibinfo {author} {\bibfnamefont {M.}~\bibnamefont
  {Sadiq}}, \bibinfo {author} {\bibfnamefont {P.}~\bibnamefont {Badzi{\k a}g}},
  \bibinfo {author} {\bibfnamefont {M.}~\bibnamefont {Bourennane}},\ and\
  \bibinfo {author} {\bibfnamefont {A.}~\bibnamefont {Cabello}},\ }\href
  {https://doi.org/10.1103/PhysRevA.87.012128} {\bibfield  {journal} {\bibinfo
  {journal} {Phys. Rev. A}\ }\textbf {\bibinfo {volume} {87}},\ \bibinfo
  {pages} {012128} (\bibinfo {year} {2013})}\BibitemShut {NoStop}%
\bibitem [{\citenamefont {Cabello}\ \emph {et~al.}(2010)\citenamefont
  {Cabello}, \citenamefont {Severini},\ and\ \citenamefont {Winter}}]{CSW1}%
  \BibitemOpen
  \bibfield  {author} {\bibinfo {author} {\bibfnamefont {A.}~\bibnamefont
  {Cabello}}, \bibinfo {author} {\bibfnamefont {S.}~\bibnamefont {Severini}},\
  and\ \bibinfo {author} {\bibfnamefont {A.}~\bibnamefont {Winter}},\
  }\href@noop {} {\bibinfo {title} {({N}on-)contextuality of physical theories
  as an axiom.}} (\bibinfo {year} {2010}),\ \Eprint
  {https://arxiv.org/abs/1010.2163} {arXiv:1010.2163} \BibitemShut {NoStop}%
\bibitem [{\citenamefont {Cabello}\ \emph {et~al.}(2014)\citenamefont
  {Cabello}, \citenamefont {Severini},\ and\ \citenamefont {Winter}}]{CSW2}%
  \BibitemOpen
  \bibfield  {author} {\bibinfo {author} {\bibfnamefont {A.}~\bibnamefont
  {Cabello}}, \bibinfo {author} {\bibfnamefont {S.}~\bibnamefont {Severini}},\
  and\ \bibinfo {author} {\bibfnamefont {A.}~\bibnamefont {Winter}},\ }\href
  {https://doi.org/10.1103/PhysRevLett.112.040401} {\bibfield  {journal}
  {\bibinfo  {journal} {Phys. Rev. Lett.}\ }\textbf {\bibinfo {volume} {112}},\
  \bibinfo {pages} {040401} (\bibinfo {year} {2014})}\BibitemShut {NoStop}%
\bibitem [{\citenamefont {Lov\'asz}(1979)}]{Lovasz}%
  \BibitemOpen
  \bibfield  {author} {\bibinfo {author} {\bibfnamefont {L.}~\bibnamefont
  {Lov\'asz}},\ }\href {https://doi.org/10.1109/TIT.1979.1055985} {\bibfield
  {journal} {\bibinfo  {journal} {IEEE Trans. Inf. Theory}\ }\textbf {\bibinfo
  {volume} {25}},\ \bibinfo {pages} {1} (\bibinfo {year} {1979})}\BibitemShut
  {NoStop}%
\bibitem [{\citenamefont {Cirel'son}(1980)}]{Tsirelson}%
  \BibitemOpen
  \bibfield  {author} {\bibinfo {author} {\bibfnamefont {B.~S.}\ \bibnamefont
  {Cirel'son}},\ }\href {https://doi.org/10.1007/BF00417500} {\bibfield
  {journal} {\bibinfo  {journal} {Lett. Math. Phys.}\ }\textbf {\bibinfo
  {volume} {4}},\ \bibinfo {pages} {93} (\bibinfo {year} {1980})}\BibitemShut
  {NoStop}%
\bibitem [{\citenamefont {Knuth}(1994)}]{sandwich}%
  \BibitemOpen
  \bibfield  {author} {\bibinfo {author} {\bibfnamefont {D.~E.}\ \bibnamefont
  {Knuth}},\ }\href {https://doi.org/https://doi.org/10.37236/1193} {\bibfield
  {journal} {\bibinfo  {journal} {Electron. J. Comb.}\ }\textbf {\bibinfo
  {volume} {1}},\ \bibinfo {pages} {A1} (\bibinfo {year} {1994})}\BibitemShut
  {NoStop}%
\bibitem [{\citenamefont {Rabelo}\ \emph {et~al.}(2014)\citenamefont {Rabelo},
  \citenamefont {Duarte}, \citenamefont {L{\'{o}}pez-Tarrida}, \citenamefont
  {{Terra Cunha}},\ and\ \citenamefont {Cabello}}]{MultGraphAppr}%
  \BibitemOpen
  \bibfield  {author} {\bibinfo {author} {\bibfnamefont {R.}~\bibnamefont
  {Rabelo}}, \bibinfo {author} {\bibfnamefont {C.}~\bibnamefont {Duarte}},
  \bibinfo {author} {\bibfnamefont {A.~J.}\ \bibnamefont
  {L{\'{o}}pez-Tarrida}}, \bibinfo {author} {\bibfnamefont {M.}~\bibnamefont
  {{Terra Cunha}}},\ and\ \bibinfo {author} {\bibfnamefont {A.}~\bibnamefont
  {Cabello}},\ }\href {https://doi.org/10.1088/1751-8113/47/42/424021}
  {\bibfield  {journal} {\bibinfo  {journal} {J. Phys. A Math. Theor.}\
  }\textbf {\bibinfo {volume} {47}},\ \bibinfo {pages} {424021} (\bibinfo
  {year} {2014})}\BibitemShut {NoStop}%
\bibitem [{\citenamefont {Le}\ \emph {et~al.}(2021)\citenamefont {Le},
  \citenamefont {Meroni}, \citenamefont {Sturmfels}, \citenamefont {Werner},\
  and\ \citenamefont {Ziegler}}]{Le22}%
  \BibitemOpen
  \bibfield  {author} {\bibinfo {author} {\bibfnamefont {T.~P.}\ \bibnamefont
  {Le}}, \bibinfo {author} {\bibfnamefont {C.}~\bibnamefont {Meroni}}, \bibinfo
  {author} {\bibfnamefont {B.}~\bibnamefont {Sturmfels}}, \bibinfo {author}
  {\bibfnamefont {R.~F.}\ \bibnamefont {Werner}},\ and\ \bibinfo {author}
  {\bibfnamefont {T.}~\bibnamefont {Ziegler}},\ }\href
  {https://doi.org/10.48550/arXiv.2111.06270} {\bibinfo {title} {Quantum
  correlations in the minimal scenario}} (\bibinfo {year} {2021}),\ \Eprint
  {https://arxiv.org/abs/2111.06270} {arXiv:2111.06270} \BibitemShut {NoStop}%
\bibitem [{\citenamefont {Vandr\'e}(2020)}]{MA}%
  \BibitemOpen
  \bibfield  {author} {\bibinfo {author} {\bibfnamefont {L.}~\bibnamefont
  {Vandr\'e}},\ }\emph {\bibinfo {title}
  {\href{https://diglib.uibk.ac.at/urn:nbn:at:at-ubi:1-72389}{On Quantum Sets
  of Non-Contextuality Inequalities}}},\ \href
  {https://diglib.uibk.ac.at/urn:nbn:at:at-ubi:1-72389} {\bibinfo {type}
  {{Master's Thesis}}},\ \bibinfo  {school} {University of Innsbruck} (\bibinfo
  {year} {2020})\BibitemShut {NoStop}%
\bibitem [{\citenamefont {Amaral}\ and\ \citenamefont {{Terra
  Cunha}}(2018)}]{GraphBook}%
  \BibitemOpen
  \bibfield  {author} {\bibinfo {author} {\bibfnamefont {B.}~\bibnamefont
  {Amaral}}\ and\ \bibinfo {author} {\bibfnamefont {M.}~\bibnamefont {{Terra
  Cunha}}},\ }\href@noop {} {\emph {\bibinfo {title}
  {\href{https://doi.org/10.1007/978-3-319-93827-1}{On Graph Approaches to
  Contextuality and their Role in Quantum Theory}}}}\ (\bibinfo  {publisher}
  {Springer International Publishing},\ \bibinfo {year} {2018})\BibitemShut
  {NoStop}%
\bibitem [{\citenamefont {Budroni}\ \emph {et~al.}(2022)\citenamefont
  {Budroni}, \citenamefont {Cabello}, \citenamefont {G\"uhne}, \citenamefont
  {Kleinmann},\ and\ \citenamefont {Larsson}}]{ContexReview}%
  \BibitemOpen
  \bibfield  {author} {\bibinfo {author} {\bibfnamefont {C.}~\bibnamefont
  {Budroni}}, \bibinfo {author} {\bibfnamefont {A.}~\bibnamefont {Cabello}},
  \bibinfo {author} {\bibfnamefont {O.}~\bibnamefont {G\"uhne}}, \bibinfo
  {author} {\bibfnamefont {M.}~\bibnamefont {Kleinmann}},\ and\ \bibinfo
  {author} {\bibfnamefont {J.-A.}\ \bibnamefont {Larsson}},\ }\href
  {https://doi.org/10.1103/RevModPhys.94.045007} {\bibfield  {journal}
  {\bibinfo  {journal} {Rev. Mod. Phys.}\ }\textbf {\bibinfo {volume} {94}},\
  \bibinfo {pages} {045007} (\bibinfo {year} {2022})}\BibitemShut {NoStop}%
\bibitem [{\citenamefont {Lovász}(1994)}]{LOVASZ1994137}%
  \BibitemOpen
  \bibfield  {author} {\bibinfo {author} {\bibfnamefont {L.}~\bibnamefont
  {Lovász}},\ }\href
  {https://doi.org/https://doi.org/10.1016/0012-365X(92)00057-X} {\bibfield
  {journal} {\bibinfo  {journal} {Discrete Math.}\ }\textbf {\bibinfo {volume}
  {124}},\ \bibinfo {pages} {137} (\bibinfo {year} {1994})}\BibitemShut
  {NoStop}%
\bibitem [{\citenamefont {Slofstra}(2019)}]{slofstra_2019}%
  \BibitemOpen
  \bibfield  {author} {\bibinfo {author} {\bibfnamefont {W.}~\bibnamefont
  {Slofstra}},\ }\href {https://doi.org/10.1017/fmp.2018.3} {\bibfield
  {journal} {\bibinfo  {journal} {Forum Math. Pi}\ }\textbf {\bibinfo {volume}
  {7}},\ \bibinfo {pages} {e1} (\bibinfo {year} {2019})}\BibitemShut {NoStop}%
\bibitem [{\citenamefont {{Abu Ashik M.\ Irfan}}\ \emph
  {et~al.}(2020)\citenamefont {{Abu Ashik M.\ Irfan}}, \citenamefont {Mayer},
  \citenamefont {Ortiz},\ and\ \citenamefont {Knill}}]{Irfan20}%
  \BibitemOpen
  \bibfield  {author} {\bibinfo {author} {\bibnamefont {{Abu Ashik M.\
  Irfan}}}, \bibinfo {author} {\bibfnamefont {K.}~\bibnamefont {Mayer}},
  \bibinfo {author} {\bibfnamefont {G.}~\bibnamefont {Ortiz}},\ and\ \bibinfo
  {author} {\bibfnamefont {E.}~\bibnamefont {Knill}},\ }\href
  {https://doi.org/10.1103/PhysRevA.101.032106} {\bibfield  {journal} {\bibinfo
   {journal} {Phys. Rev. A}\ }\textbf {\bibinfo {volume} {101}},\ \bibinfo
  {pages} {032106} (\bibinfo {year} {2020})}\BibitemShut {NoStop}%
\bibitem [{\citenamefont {Peres}(1990)}]{peres1990neumark}%
  \BibitemOpen
  \bibfield  {author} {\bibinfo {author} {\bibfnamefont {A.}~\bibnamefont
  {Peres}},\ }\href {https://doi.org/https://doi.org/10.1007/BF01883517}
  {\bibfield  {journal} {\bibinfo  {journal} {Found. Phys.}\ }\textbf {\bibinfo
  {volume} {20}},\ \bibinfo {pages} {1441} (\bibinfo {year}
  {1990})}\BibitemShut {NoStop}%
\bibitem [{\citenamefont {Nielsen}\ and\ \citenamefont
  {Chuang}(2010)}]{NielsenChuang}%
  \BibitemOpen
  \bibfield  {author} {\bibinfo {author} {\bibfnamefont {M.~A.}\ \bibnamefont
  {Nielsen}}\ and\ \bibinfo {author} {\bibfnamefont {I.~L.}\ \bibnamefont
  {Chuang}},\ }\href@noop {} {\emph {\bibinfo {title}
  {\href{https://doi.org/10.1017/CBO9780511976667}{Quantum Computation and
  Quantum Information}}}}\ (\bibinfo  {publisher} {Cambridge University
  Press},\ \bibinfo {year} {2010})\BibitemShut {NoStop}%
\bibitem [{\citenamefont {P\'al}\ and\ \citenamefont
  {V\'ertesi}(2009)}]{concavity}%
  \BibitemOpen
  \bibfield  {author} {\bibinfo {author} {\bibfnamefont {K.~F.}\ \bibnamefont
  {P\'al}}\ and\ \bibinfo {author} {\bibfnamefont {T.}~\bibnamefont
  {V\'ertesi}},\ }\href {https://doi.org/10.1103/physreva.80.042114} {\bibfield
   {journal} {\bibinfo  {journal} {Phys. Rev. A}\ }\textbf {\bibinfo {volume}
  {80}},\ \bibinfo {pages} {042114} (\bibinfo {year} {2009})}\BibitemShut
  {NoStop}%
\bibitem [{\citenamefont {Navascu\'es}\ \emph {et~al.}(2007)\citenamefont
  {Navascu\'es}, \citenamefont {Pironio},\ and\ \citenamefont
  {Ac\'{\i}n}}]{NPA07}%
  \BibitemOpen
  \bibfield  {author} {\bibinfo {author} {\bibfnamefont {M.}~\bibnamefont
  {Navascu\'es}}, \bibinfo {author} {\bibfnamefont {S.}~\bibnamefont
  {Pironio}},\ and\ \bibinfo {author} {\bibfnamefont {A.}~\bibnamefont
  {Ac\'{\i}n}},\ }\href {https://doi.org/10.1103/PhysRevLett.98.010401}
  {\bibfield  {journal} {\bibinfo  {journal} {Phys. Rev. Lett.}\ }\textbf
  {\bibinfo {volume} {98}},\ \bibinfo {pages} {010401} (\bibinfo {year}
  {2007})}\BibitemShut {NoStop}%
\bibitem [{\citenamefont {Navascu{\'{e}}s}\ \emph {et~al.}(2008)\citenamefont
  {Navascu{\'{e}}s}, \citenamefont {Pironio},\ and\ \citenamefont
  {Ac{\'{\i}}n}}]{NPA08}%
  \BibitemOpen
  \bibfield  {author} {\bibinfo {author} {\bibfnamefont {M.}~\bibnamefont
  {Navascu{\'{e}}s}}, \bibinfo {author} {\bibfnamefont {S.}~\bibnamefont
  {Pironio}},\ and\ \bibinfo {author} {\bibfnamefont {A.}~\bibnamefont
  {Ac{\'{\i}}n}},\ }\href {https://doi.org/10.1088/1367-2630/10/7/073013}
  {\bibfield  {journal} {\bibinfo  {journal} {New J. Phys.}\ }\textbf {\bibinfo
  {volume} {10}},\ \bibinfo {pages} {073013} (\bibinfo {year}
  {2008})}\BibitemShut {NoStop}%
\bibitem [{\citenamefont {{Terra Cunha}}(2021)}]{building_blocks}%
  \BibitemOpen
  \bibfield  {author} {\bibinfo {author} {\bibfnamefont {M.}~\bibnamefont
  {{Terra Cunha}}},\ }\href {https://www.youtube.com/watch?v=CI0sidUkcyI}
  {\bibinfo {title}
  {\href{https://www.youtube.com/watch?v=CI0sidUkcyI}{{C}oloured Graph Approach
  for contextuality with parts.} {QCQMB C}olloquium}} (\bibinfo {year}
  {2021})\BibitemShut {NoStop}%
\bibitem [{\citenamefont {Vandr\'e}\ and\ \citenamefont {{Terra
  Cunha}}(2021)}]{supplement}%
  \BibitemOpen
  \bibfield  {author} {\bibinfo {author} {\bibfnamefont {L.}~\bibnamefont
  {Vandr\'e}}\ and\ \bibinfo {author} {\bibfnamefont {M.}~\bibnamefont {{Terra
  Cunha}}},\ }\href
  {https://gitlab.com/alvla/masterarbeit/-/blob/master/Quantum_sets_LV.py}
  {\bibinfo {title} {Supplementary material.
  \url{https://gitlab.com/alvla/supplemental_quantum_sets}}} (\bibinfo {year}
  {2021})\BibitemShut {NoStop}%
\bibitem [{\citenamefont {P{\'{a}}l}\ and\ \citenamefont
  {V{\'{e}}rtesi}(2010)}]{Pal10}%
  \BibitemOpen
  \bibfield  {author} {\bibinfo {author} {\bibfnamefont {K.~F.}\ \bibnamefont
  {P{\'{a}}l}}\ and\ \bibinfo {author} {\bibfnamefont {T.}~\bibnamefont
  {V{\'{e}}rtesi}},\ }\href {https://doi.org/10.1103/physreva.82.022116}
  {\bibfield  {journal} {\bibinfo  {journal} {Phys. Rev. A}\ }\textbf {\bibinfo
  {volume} {82}},\ \bibinfo {pages} {022116} (\bibinfo {year}
  {2010})}\BibitemShut {NoStop}%
\bibitem [{\citenamefont {Ray}\ \emph {et~al.}(2021)\citenamefont {Ray},
  \citenamefont {Boddu}, \citenamefont {Bharti}, \citenamefont {Kwek},\ and\
  \citenamefont {Cabello}}]{dimension_wittnessing}%
  \BibitemOpen
  \bibfield  {author} {\bibinfo {author} {\bibfnamefont {M.}~\bibnamefont
  {Ray}}, \bibinfo {author} {\bibfnamefont {N.~G.}\ \bibnamefont {Boddu}},
  \bibinfo {author} {\bibfnamefont {K.}~\bibnamefont {Bharti}}, \bibinfo
  {author} {\bibfnamefont {L.-C.}\ \bibnamefont {Kwek}},\ and\ \bibinfo
  {author} {\bibfnamefont {A.}~\bibnamefont {Cabello}},\ }\href
  {https://doi.org/10.1088/1367-2630/abcacd} {\bibfield  {journal} {\bibinfo
  {journal} {New J. Phys.}\ }\textbf {\bibinfo {volume} {23}},\ \bibinfo
  {pages} {033006} (\bibinfo {year} {2021})}\BibitemShut {NoStop}%
\bibitem [{\citenamefont {Wittek}(2015)}]{ncpol2sdpa}%
  \BibitemOpen
  \bibfield  {author} {\bibinfo {author} {\bibfnamefont {P.}~\bibnamefont
  {Wittek}},\ }\href {https://doi.org/10.1145/2699464} {\bibfield  {journal}
  {\bibinfo  {journal} {ACM Trans. Math. Softw.}\ }\textbf {\bibinfo {volume}
  {41}},\ \bibinfo {pages} {1} (\bibinfo {year} {2015})}\BibitemShut {NoStop}%
\bibitem [{\citenamefont {ApS}(2020)}]{mosek}%
  \BibitemOpen
  \bibfield  {author} {\bibinfo {author} {\bibfnamefont {M.}~\bibnamefont
  {ApS}},\ }\href {https://docs.mosek.com/9.2/pythonapi/index.html} {\bibinfo
  {title} {\href{https://docs.mosek.com/9.2/pythonapi/index.html}{MOSEK
  Optimizer API for Python 9.2}}} (\bibinfo {year} {2020})\BibitemShut
  {NoStop}%
\bibitem [{\citenamefont {Schmidt}(1907)}]{Schmidt}%
  \BibitemOpen
  \bibfield  {author} {\bibinfo {author} {\bibfnamefont {E.}~\bibnamefont
  {Schmidt}},\ }\href {https://doi.org/10.1007/BF01449770} {\bibfield
  {journal} {\bibinfo  {journal} {Math. Ann.}\ }\textbf {\bibinfo {volume}
  {63}},\ \bibinfo {pages} {433–476} (\bibinfo {year} {1907})}\BibitemShut
  {NoStop}%
\end{thebibliography}%

\onecolumngrid
\appendix

\section{Analytical Calculations of the Behaviors}\label{Sapp_detail}

In this appendix we want to give an idea of how the handles and projectors which lead to quantum behaviors on the boundaries of the required quantum sets can be constructed. 
A detailed description 
can be found in Ref.\ \cite{MA}. 

\begin{figure}[h]
\centering
\includegraphics[scale=1]{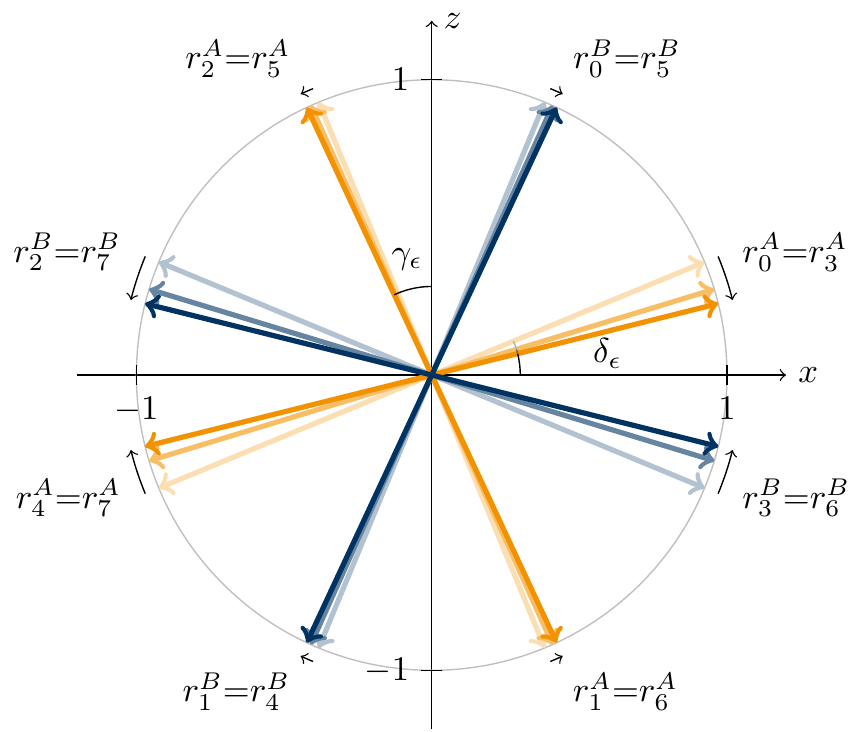}
\caption{Bloch vectors of the orthogonal labeling and the states which lead to the colored Lov\'asz numbers $\vartheta_c\left(\Gchsh, \omega_5^{\varepsilon = 0}(0^-1^-7^-) \right)$ and $\vartheta_c\left(\Gchsh, \omega_5^{\varepsilon = 1}(0^-1^-7^-) \right)$ are known. The intermediate colored Lov\'asz numbers can be calculated by a transition from one case to the other. 
The players are also indicated by the superscript $A$ or $B$. For each family of vectors, the light arrows are the Bloch vectors for the case $\varepsilon = 0$ (CHSH) while the dark arrows are the Bloch vectors for the case $\varepsilon = 1$ ($I^P_1$). The intermediate vectors and shared states are given in Eqs.\ \eqref{E_a_chsh_pent}-\eqref{E_delt_chsh_pent}.} \label{Fblochchshrotpenta}
\end{figure}

If we restrict Alice and Bob to useing measurements and states acting on a Hilbert space $\HH^2 \otimes \HH^2$,
the projectors of the orthogonal labeling can be represented in the Pauli basis:
\begin{subequations}
\begin{align}
\Pi &= \Pi^A \otimes \Pi^B \\
&= \frac{1}{2} \left(\1 + \vec{r}^A \vec{\sigma} \right) \otimes \frac{1}{2} \left(\1 + \vec{r}^B \vec{\sigma} \right), \label{EPi_bloch}
\end{align} 
\end{subequations}
where $\vec{r}^j$ is a Bloch vector and $\vec{\sigma}$ is a vector with the Pauli matrices $\sigma_x, \sigma_y$, and $\sigma_z$ as components. 
The Bloch vectors can be represented on a Bloch sphere (see {e.g.}\ Ref.\ \cite{NielsenChuang}).
In the following we will work with an equator of the Bloch sphere for which $r^j_y = 0$.
Due to the Schmidt decomposition \cite{Schmidt} and up to local operations, a bipartite entangled state on $\HH^2 \otimes \HH^2$ can be written as $\ket{\Psi} = a \ket{00} + b \ket{11}$, with $ a^2 + b^2 = 1$.
The expectation value of a projector in the form of \cref{EPi_bloch} due to this state
is given by the equation:
\begin{align}
\expval{\Pi}{\Psi} &= \frac{1}{4} \left[ 1 + 2 a b  r_{x}^A  r_{x}^B + r_{z}^A  r_{z}^B + (2a^2 -1) \left(r_{z}^A + r_{z}^B \right) \right]. \label{Eexpallgem}
\end{align}
Note that if $\ket{\Psi}$ is a maximally entangled state, i.e., $a=b=\nicefrac{1}{\sq}$, \cref{Eexpallgem} reduces to $\nicefrac{1 + \vec{r}^A \cdot \vec{r}^B}{4}$.
On the other hand, for non maximally entangled states, it is relevant in which quadrants the vectors are.
The maximal quantum bound of the CHSH inequality can be reached with a maximally entangled state and projectors in the form of \cref{EPi_bloch} with Bloch vectors as shown in \cref{Fblochchshrotpenta} by transparent orange (light) and blue (dark) vectors. We see that the vectors are distributed in a very symmetric way.
For other inequalities, the maximally quantum bound can be reached by non-maximally entangled states and a less symmetric distribution of Bloch vectors. This is the case for the less symmetric graph of the first pentagonal inequality $I^P_1$. The maximally quantum bound of $I^P_1$ is reached by an entangled state $\ket{\Psi}= a_P \ket{00} + b_P \ket{11}$ with $a_P \approx 0.6338$, $b_P \approx 0.7735$  and projectors in the form of \cref{EPi_bloch} with Bloch vectors as shown in \cref{Fblochchshrotpenta} by solid orange (light) and blue vectors. The angles are given by $\gamma_{P_1} \approx  25\degree$ and $\delta_{P_1} \approx  14\degree$ \cite{PentBI}.

\begin{figure}
\centering
\includegraphics[scale=1]{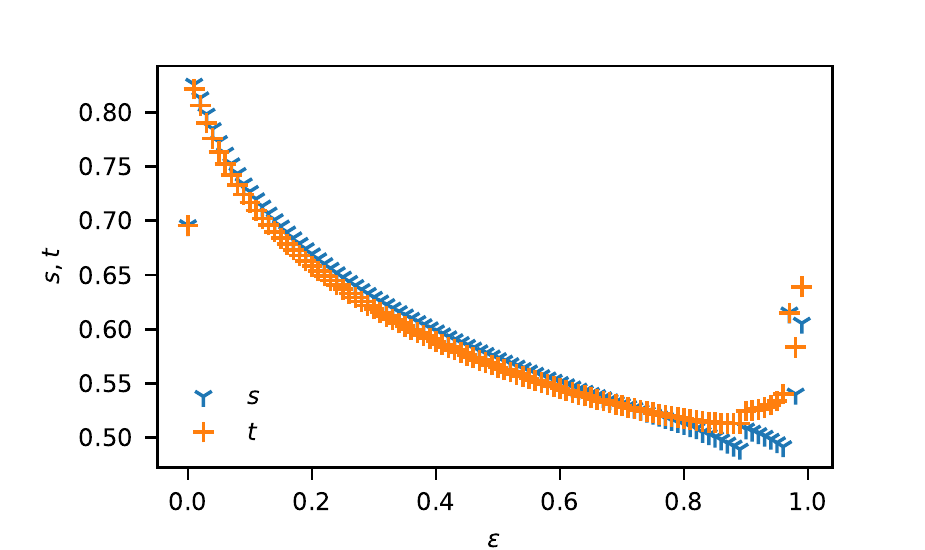}
\caption{For the parametrization in Eqs.\ \eqref{E_a_chsh_pent}-\eqref{E_delt_chsh_pent} we computed the parameters $s$ and $t$ shown in this figure to be optimal. Caused by the irregular pattern of the parameters, we conjecture that there is a better parametrization.} \label{Fparameter-s-t}
\end{figure}

We computed the maximally quantum bound of the set of equations $S(\GG,{\omega}) = \sum_{i = 0}^{n-1} \omega_i  P_i$ with a weight vector $\omega$ as given in \cref{Eomega_5} and plotted it in \cref{FG_5_compare}. For $\varepsilon = 0$ and $1$ we have the known CHSH and $I^P_1$ inequalities. The intermediate cases can be computed by a superposition of both handles and projectors where the Bloch vectors rotate from one case to the other. 
The handle is given by $\ket{\Psi} = a_{\varepsilon} \ket{00} + b_{\varepsilon} \ket{11}$ with 
\begin{subequations}
\begin{align}
a_{\varepsilon} &= \left( 1- \varepsilon^s \right) \frac{1}{\sq} + \varepsilon^s 0.6338, \label{E_a_chsh_pent} \\
b_{\varepsilon} &= \sqrt{1-a_{\varepsilon}^2}, \label{E_b_chsh_pent}
\end{align}
\end{subequations}
and the angles of Bloch vectors in \cref{Fblochchshrotpenta} are given by
\begin{subequations}
\begin{align}
\gamma_{\varepsilon} &= \left( 1- \varepsilon^t \right) \frac{\pi}{8} + \varepsilon^t \gamma_{P_1}, \\
\delta_{\varepsilon} &= \left( 1- \varepsilon^t \right) \frac{\pi}{8} + \varepsilon^t \delta_{P_1}. \label{E_delt_chsh_pent}
\end{align}
\end{subequations}
Choosing fixed parameters $t$ and $s$ in the range $0.6 \lesssim s, t \lesssim 0.67$, we get lower bounds to quantum violations which deviate from the upper bound of order $\OO(10^{-5})$.
Slightly better values can be reached by the parameters shown in \cref{Fparameter-s-t}. The concrete $\varepsilon$ dependence of $t$ and $s$  is not found yet. We conjecture that one can find a better parametrization than the one we proposed here.

The curve for $\Gvier$ was constructed with
orthogonal labelings and states acting on a Hilbert space $\HH^2 \otimes \HH^2$ as well. We saw that using this dimensions, we can construct behaviors in $\Qvier$ which do not belong to $\Qchsh$, even if we did not find the optimal behaviors for all $\varepsilon$. We conjecture that the maximal quantum bound can be reached in this dimension. 
In the graph $\Gvier$, there are two nonadjacent subgraphs of Alice but four of Bob. Therefore, there are already more options of arranging the Bloch vectors in a Hilbert space $\HH^2 \otimes \HH^2$ than we had for $\Gchsh$.

\begin{figure}
\centering
\includegraphics[scale=1]{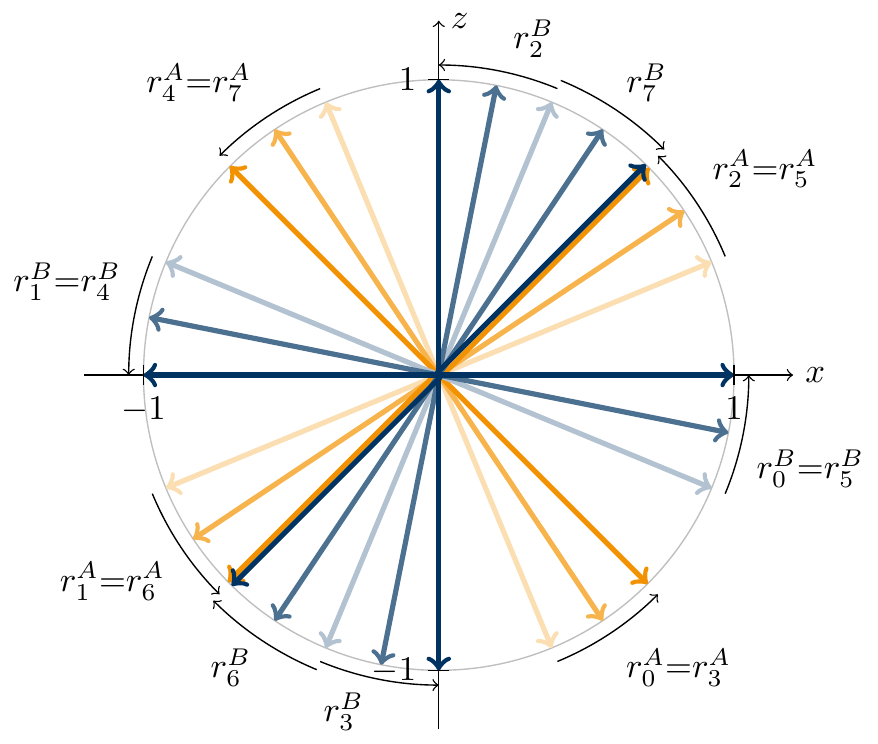}
\caption{Ansatz to find the behaviors which lead to the colored Lov\'asz number $\vartheta_c\left(\Gvier, \omega_5^{\varepsilon}(0^-1^-7^-) \right)$. We rotate $r_6^B$ and $r_7^B$ counterclockwise and all other vectors clockwise with the angle $\gamma = \varepsilon^t \frac{\pi}{8}$, where $t \approx 1$. The transparent orange (light) and blue (dark) vectors represent the vectors for Alice and Bob at $\varepsilon = 0$, while the dark vectors lead to the maximal quantum bound at $\varepsilon = 1$.} \label{FblochG441111rot}
\end{figure}

As we did for the CHSH inequality, we explored the set of functions $S(\GG,{\omega}) = \sum_{i = 0}^{n-1} \omega_i P_i$ with the same weight vector $\omega$. 
In the case of $\varepsilon = 0$, the maximum quantum value is the same as for the CHSH inequality and the handle as well as the set of projectors is known.
For $\varepsilon = 1$, we have the also known third pentagonal inequality $I^P_3$ \cite{PentBI}. Their quantum bounds can be reached by the maximally entangled Bell state $\ket{\Psi^+}$ and the Bloch vectors as shown in \cref{FblochG441111rot} in solid orange (light) and blue (dark).
As an ansatz to find the behaviors for $0 < \varepsilon < 1$ 
we rotate the vectors as shown in \cref{FblochG441111rot}.
We also vary the parameter $a$ by the factor $(1-s)$, where the minimum of $s$ is $s=0$ for $\varepsilon = 0$ and $1$ and the maximum is $s \approx 0.027$ for $\varepsilon \approx 0.5$.

We next discuss the curve of $\Gdrei$. 
In the same way as for $\Gchsh$, for $\Gdrei$ there are two nonadjacent subgraphs for each color. 
If we restrict the Hilbert space dimension to $\HH^2 \otimes \HH^2$, the options of arranging the Bloch vectors are the same as for $\Gchsh$.
We need to allow at least one party to use three dimensions in order to have more options in the orthogonal labeling. Note that more options in the orthogonal labeling does not necessarily lead to new behaviors $P$.

In \cref{FG_5_compare}
we see that for $\varepsilon \lesssim 0.85$ the colored Lov\'asz number $\vartheta_c\left(\Gdrei, \omega_5^{\varepsilon}(0^-1^-7^-) \right)$ is the same as for $\vartheta_c\left(\Gchsh, \omega_5^{\varepsilon}(0^-1^-7^-) \right)$. At $\varepsilon \approx 0.85$
we have a kink which is a discontinuity in the first derivative and for $\varepsilon \gtrsim 0.85$ we see that 
$ \vartheta_c\left(\Gdrei, \omega_5^{\varepsilon}(0^-1^-7^-) \right)$ becomes greater than $\vartheta_c\left(\Gchsh, \omega_5^{\varepsilon}(0^-1^-7^-) \right)$ and matches with the curve of $\vartheta_c\left(\Gvier, \omega_5^{\varepsilon}(0^-1^-7^-) \right)$ at $\varepsilon = 1$. The parts in which the first derivative is continuous can be explained with a smooth change of parameters as we have seen for $\Gchsh$. A kink, as we have at $\varepsilon \approx 0.85$, indicates that there are different configurations of orthogonal labelings and states
giving the same result at the kink.
Depending on $\varepsilon$, one or the other leads to a higher value of $\vartheta_c$ on different sides of the kink.

As stated before, for orthogonal labelings and states restricted to $\HH^2 \otimes \HH^2$, the set of behaviors of $\Gdrei$ is the same as the set of behaviors of $\Gchsh$. Therefore, we need a higher dimensional Hilbert space to find the behaviors which lead to the colored Lov\'asz number in the domain $\varepsilon \gtrsim 0.85$. We first explain how to get the behavior for $\varepsilon = 1$. The orthogonal labeling and the state we found act on the Hilbert space $\HH^3 \otimes \HH^2$. 
We embed the state $\ket{\Psi^+} = \frac{1}{\sq} \left( \ket{00} + \ket{11} \right)$ canonically in $\HH^3 \otimes \HH^2$, as well as the projectors induced by Bloch vectors $r_1^B$, $r_i^A$, and $r_i^B$ for $i = 0, 2, \dots, 7$. We choose the projector $\Pi_1^A = \dyad{2}$. 
Note that already $P_{33,33}(\omega_5^{\varepsilon=1}(0^-1^-7^-)) \notin \Qchsh$, which proves that $\Qdrei$ is strictly larger than $\Qchsh$.
The curve in the domain $0.85 \lesssim \varepsilon \leqslant 1$ can be constructed by projectors $\Pi_1^A = \dyad{2}$ and projectors induced by Bloch vectors $r_1^B$, $r_i^A$, and $r_i^B$ for $i = 0, 2, \dots, 7$ which rotate from the positions given above for $\Gchsh$ for $\varepsilon \approx 0.85$ to the positions which were shown to be optimal for $I_3^P$. We use a shared state which is a superposition of the state given above for $\varepsilon \approx 0.85$ and $\ket{\Psi^+}$. It is interesting to see that the behavior we found which gave us the maximal quantum bound has one component equal to 0 and all other components are computed from states and projectors in $\HH^2 \otimes \HH^2$.

{
\section{{Making Use of Higher Hilbert Space Dimensions}}\label{Sapp_dilation}

In this appendix we comment in more detail on the statement about higher Hilbert space dimensions we made earlier in the paper. We mainly summarize two references and state how to combine dilation of measurements and purification of states.
In the case of positive-operator-valued measures (POVMs), we use Naimark's theorem \cite{peres1990neumark} as presented in Ref.~\cite{Irfan20}. 
 In the case of states, we refer to the work of Nielsen and Chuang \cite{NielsenChuang}. 
 As we are dealing with finite numbers of measurements and outcomes, we do not consider the infinite case here.

\subsection*{Dilation from POVMs to PVMs}
We first state results from Appendix B of Ref.\ \cite{Irfan20} and comment on how to adapt them to our case.

Given a POVM $\lbrace \pi_i \rbrace_{i=0}^{d-1}$ acting on system $S$, there exists a projective measurement $\lbrace \Pi_i \rbrace_{i=0}^{d-1}$ on a system $S \otimes E$ where the dimension of $E$ is $d$ such that 
\begin{align}
\tr{\pi_i \rho^S} = \tr{\Pi_i (\rho^S \otimes \dyad{i}_E)}.
\end{align}
This was first proven in Ref.\ \cite{peres1990neumark}.
The projective measurement operators are constructed by
\begin{align}
\Pi_i = U^{\dagger}_{S,E} (\1_S \otimes \dyad{0}_E ) U_{S,E},
\end{align} 
where $U_{S,E}$ is a unitary on the extended system and $ \{\ket{i}\} $ is a  basis for $E$.
We want the new set of projectors to have the same pairwise commutativity structure as the initial POVM. In Ref.\ \cite{Irfan20} it was proven for two-outcome measurements that this condition is fulfilled by constructing $U_{S,E}$ such that 
\begin{align}
U_{S,E} \ket{\Psi}_S \otimes \ket{i}_E = \sum_{j=0}^{1} (-1)^{i j} \sqrt{\pi_{i \oplus j}}
\ket{\Psi}_S \otimes \ket{j}_E.
\end{align}
Note, that it was proven for the special case $d=2$. In our case, this is sufficient. 
It is not immediate to generalize this result and it would be nice to see such a construction.

In our case, we are dealing with tensor products of measurement operators: $\pi^A \otimes \pi^B$. 
In the case of multiple POVMs, Irfan \textit{et al.} propose to extend to several systems $\bigotimes_J E^J$. 
In our case, we can extend our system $S^A \otimes S^B$ to  $(S^A \otimes E^A) \otimes (S^B \otimes E^B)$.

\subsection*{Purification of states}
Purification of states follows a similar idea as presented before. In this section, we follow Ref.~\cite{NielsenChuang}.

Given a state $\rho^S$ of a quantum system $S$, it is possible to introduce a system $R$ and define a state $\ket{SR}$ such that 
\begin{equation}
\rho^S = \tr_R \left(\dyad{SR} \right). \label{E_traceR}
\end{equation}
A system $R$ and a state $\ket{SR}$ which fulfills \cref{E_traceR} can be constructed in the following way: Knowing the orthogonal decomposition 
\begin{equation}
\rho^S = \sum_n  p_n \dyad{n^S},
\end{equation}
we can define the state
\begin{equation}
\ket{SR} = \sum_n \sqrt{p_n} \ket{n^S} \ket{n^R}, \label{E_ketAR}
\end{equation}
where system $R$ has a state space isomorphic to that of system $S$, with orthonormal basis states $\lbrace \ket{n^R} \rbrace$. 
It follows that \cref{E_ketAR} fulfills the condition \eqref{E_traceR}.

We show that for a given POVM $\lbrace \pi_i \rbrace$ and a state $\rho^S$,
\begin{equation}
\tr{\pi \rho^S} = \tr{(\pi \otimes \1_R) \dyad{SR}}
\end{equation}
holds:
\begin{subequations}
\begin{align}
\tr{(\pi \otimes \1_R) \dyad{SR}} &= \tr{(\pi \otimes \1_R) (\sum_{n,m}  \sqrt{p_n p_m} \ketbra{n^S}{m^S} \otimes \ketbra{n^R}{m^R})} \\
&= \sum_{n,m} \sqrt{p_n p_m}  \tr \lbrace \pi \ketbra{n^S}{m^S} \rbrace \tr \lbrace \1_R \ketbra{n^R}{m^R} \rbrace \\
&= \sum_{n,m} \sqrt{p_n p_m}  \tr \lbrace \pi \ketbra{n^S}{m^S} \rbrace \delta_{n,m} \\
&= \sum_n p_n   \tr \lbrace \pi \ketbra{n^S}{n^S} \rbrace \\
&= \tr{\pi \rho^S}.
\end{align}
\end{subequations}

Therefore, we get the same statistics from a POVM $\lbrace \pi_i \rbrace$ and a state $\rho^S$ as we get from its purified version.

\subsection*{Combination of both methods}

As we have seen above, we can extend POVMs to PVMs and get the same statistics by only simple modifications of the state and equivalently going from mixed states to pure states by doing simple modifications of the measurement sets. In both cases, we make use of embedding the system canonically to a higher-dimensional system. We can combine the two methods by embedding the measurement sets and the state into different higher-dimensional spaces. 
Therefore, it is sufficient to consider PVMs and pure states.
Here, the POVM purification follows the locality demand by using different state spaces for each party, while the state is purified globally, since entangled states are naturally welcome in this discussion.

}

\end{document}